# Classical Structures in Quantum Mechanics and Applications

Augusto César Lobo[1] and Clyffe de Assis Ribeiro[2]

1 - Post-Doctoral Visiting Scholar

Institute for Quantum Studies-Chapman University, California, US §

2 - Physics Department

Universidade Federal de Minas Gerais, Belo Horizonte, Brazil

1 - Corresponding author e-mail: lobo@chapman.edu

§ - On leave from the Physics Department of the Federal University of Ouro Preto, Minas Gerais, Brazil

**Acknowledgments**

A. C. Lobo and C. A. Ribeiro wishtoacknowledge financial supportfrom Conselho Nacional de Desenvolvimento Científico e Tecnológico (CNPq). A. C. Lobo also thanks the Institute for Quantum Studies at Chapman University for providing the means to finish this project.

**This work is dedicated to Maria Carolina Nemes on occasion of her sixtieth birthday.**

The theory of Non-Relativistic Quantum Mechanics was created (or discovered) back in the 1920's mainly by Schrödinger and Heisenberg, but it is fair enough to say that a more modern and unified approach to the subject was introduced by Dirac and Jordan with their (intrinsic) Transformation Theory. In his famous text book on quantum mechanics [1], Dirac introduced his well-known bra and ket notation and a view that even Einstein (who was, as well known, very critical towards the general quantum physical world-view) considered the most elegant presentation of the theory at that time[2]. One characteristic of this formulation is that the observables of position and momentum are truly treated equally so that an intrinsic phase-space approach seems a natural course to be taken. In fact, we may distinguish at least two different quantum mechanical approaches to the structure of the quantum phase space: The Weyl-Wigner (WW) formalism and the advent of the theory of Coherent States (CS). The Weyl-Wigner formalism has had many applications ranging from the discussion of the Classical/Quantum Mechanical transition and quantum chaos to signal analysis[3,4]. The Coherent State formalism had a profound impact on Quantum Optics and during the course of time has found applications in diverse areas such as geometric quantization, wavelet and harmonic analysis [5]. In this chapter we present a compact review of these formalisms (with also a more intrinsic and coordinate independent notation) towards some non-standard and up-to-date applications such as modular variables and weak values. We recall that new points of view concerning a certain subject are always welcome. It is common historical knowledge that the establishment of bridges between distinct disciplines is usually a very fruitful enterprise for both subjects. This interplay has brought us (since at least from Newton and Galileo to Einstein and Minkowski, passing through Euler, Lagrange, Hamilton, Maxwell and many others) a wonderful multitude of results where mathematical structures are discovered by contemplating natures wonders and physical theories are guessed from deep and beautiful mathematics.

The structure of this chapter is the following: In section 1, we review the mathematical structure of classical phase space as the geometric notion that "classical mechanics is symplectic geometry" where one dwells with the natural symplectic structure of cotangent bundles of configuration manifolds of classical particles submitted to holonomic constraints. This symplectic geometric structure appears again and again all over the next sections. In section 2 we present an *intrinsic* formulation for the operator algebra of the WW Transform (similar to Dirac's Bra and Ket notation) that is necessary to construct this formalism in a more compact and elegant way than the usual manner. In section 3 we discuss the natural symplectic structure that projective spaces of finite dimensional Hilbert spaces inherit from their hermitian structure and for this reason are somewhat structurally analogous to classical phase spaces. In section 4, we briefly review the coherent state concept in terms of the WW basis mainly to set stage for a discussion of the quantum transforms that implement linear area preserving maps on the phase plane. In section 5, we apply some of these ideas to a phase space study of the *weak value* concept, which has become an important theoretical and experimental tool in modern investigations of quantum physics. We look at the phase space of the *measuring apparatus* that performs the weak value measurement of some quantum system. In the next section, we approach the weak measurement concept and the more general von-Neumann pre-measurement by what may be called the *opposite* approach: We look now to the geometric structure of the *measured system*. In section 7, we review Schwinger's finite Quantum Kinematics and the Finite Phase Space WW formalism. Here we present our intrinsic formalism for the Finite Weyl-Wigner Transform analogous to the continuous case developed in section 2. We also discuss how to implement a finite or discrete version of coherent states. Finally, in the last section we apply this discrete formalism to understand Aharonov's Modular Variable concept which is a key idea to understand the phenomenon of *dynamical non-locality* which plays an essential role in foundational quantum mechanical properties, from the paradigmatic quantum diffraction of particles to the subtleties of the Aharonov-Bohm effect.

# 1 – Introduction

(The main references for this section are [6] and [7])

What is meant here by a "classical structure" is the totality of the mathematical formalism associated to conservative (Hamiltonian) dynamical systems. In other words:

(i) An even-dimensional differential manifold ($\dim M = 2n$);

(ii) A canonical coordinate system $(q^i, p_i)$ $(i = 1,2, \dots, n)$ over $M$, where $n$ is the number of degrees of freedom of the system;

(iii) A non-degenerate closed 2-form $\Omega$ on $M$. In other words: $d\Omega = 0$ and the non-degeneracy means that: $\Omega(V, W) = 0 \Rightarrow V = 0$ or $W = 0, \forall\, V, W \in T_pM$ and $\forall\, p \in M$ and where $TM$ is the tangent bundle of $M$ and $T_pM$ is the tangent space (the tangent space or fiber) of $M$at $p \in M$. A differential manifold with the above structure is called *symplectic*. A theorem due to Darboux guarantees that given a point $p \in M$, there is always an open set of $M$ that contains $p$ and that admits a pair of canonical coordinates $(q^i, p_i)$ such that $\Omega = -dp_i \wedge dq^i$, where we use henceforth the sum convention unless we explicitly state the contrary. In Classical Newtonian Mechanics, the above structure arises naturally as the cotangent bundle of a configuration manifold $Q$ (of dimension $n$) of $m$ particles submitted to $3m - n$ holonomic constraints: $M = T^*Q$. In fact, any cotangent bundle has a natural 1-form $\theta = p_i dq^i$, where $q^i$ are the coordinates of $Q$ and $p_i$ are the coordinates of the convectors of $T^*Q$. It is easy to verify that indeed the 2-form defined by $\Omega = -d\theta$ satisfies automatically the conditions (i-iii). Yet not all symplectic manifolds are cotangent bundles. An example that will be of great importance for us later is that of quantum projective spaces. We may define a dynamical structure on a symplectic manifold (the phase space) by introducing a Hamiltonian function $H: M \to \mathbb{R}$ which maps a definite energy to each point of $M$. We can also define a symplectic gradient $X_H$ of $H$ (through $\Omega$) by the following relation $\Omega(X_H, Y) = dH(Y)$ for any vector field $Y$ defined over $M$. In canonical coordinates, the vector field $X_H$ is given by

$$X_H = \frac{\partial H}{\partial p_i}\frac{\partial}{\partial q^i} - \frac{\partial H}{\partial q^i}\frac{\partial}{\partial p_i} \tag{1}$$

The motion of the system is given by the integral curves of $X_H$ and the first order ODE's associated to them are the well-known Hamilton equations: $\frac{dq^i}{dt} = \frac{\partial H}{\partial p_i}$ and $\frac{dp_i}{dt} = -\frac{\partial H}{\partial q^i}$.

A classical observable is any real function defined on $M$. For example, the observables $Q^i$ and $P_i$ are respectively the i$^{\text{th}}$ component of the generalized position and momentum

$$\begin{matrix} Q^i: M \to \mathbb{R} & \quad & P^i: M \to \mathbb{R} \\ (q^i, p_i) \to q^i & \quad & (q^i, p_i) \to p^i \end{matrix} \tag{2}$$

The Poisson brackets of two observables $f$ and $g$ can then be defined as:

$$\{f, g\} = \Omega(X_f, X_g) = \frac{\partial g}{\partial p_i}\frac{\partial f}{\partial q^i} - \frac{\partial f}{\partial p_i}\frac{\partial g}{\partial q^i} \tag{3}$$

And the temporal evolution of a general observable $O(q^i, p_i)$ can be written as

$$\frac{dO}{dt} = \{O, H\} \tag{4}$$

The different formulations of quantum mechanics has shown from the very beginning a close relation to the structure of classical analytical mechanics as one can infer from common denominations of quantum physics as like, for example, the "Hamiltonian operator" for the generator of time displacements and the formal analogies of equations like the Heisenberg equation and its classical analog given by (1) (we use henceforth $\hbar = 1$ units)

$$i\dot{\hat{O}} = [\hat{O}, \hat{H}] \tag{5}$$

for the evolution of a quantum observable $\hat{O}$ and a classical observable $O(p,q)$.

Since then, some different formalisms that relate both structures have been introduced with a number of applications in physics. In this chapter, we review some of these formalisms such as the Weyl-Wigner transform theory and coherent state theory together with some up to date applications. One of the most fundamental differences between quantum physics and classical physics is that the first admits the possibilities of both discrete and continuous observables while the second admits only continuous quantities. In the next section we will discuss only continuous phase spaces and in the following sections we will address "finite" or "discrete" phase spaces. We will see that in this last case there still remains some kind of "classical skeleton" with algebraic and geometric structures that are analogous to the continuous case.

## 2 – The Weyl-Wigner Formalism

(The main references for this section are [8] and [9])

### *2.1 The Fourier Transform Operator*

Consider a quantum system defined by the motion of a single non-relativistic particle in one dimension. Let $|q(x)\rangle$ and $|p(x)\rangle$ $(-\infty < x < \infty)$ represent respectively the quantum eigenkets of the positionand momentum observables. In other words, we have:

$$\hat{Q}|q(x)\rangle = x|q(x)\rangle \quad \text{and} \quad \hat{P}|p(x)\rangle = x|p(x)\rangle \tag{6}$$

where $\hat{Q}$ and $\hat{P}$ are respectively the position and momentum operators. It is important to notice here that we use a slightly different notation than the usual choice. For instance, the ket $|q(x)\rangle$ is an eigenvector of $\hat{Q}$ with eigenvalue $x$. That is, we distinguish the "kind" of *eigenvector* (position or momentum) from its *eigenvalue*. This is different from the more common notation where $|q\rangle$ and $|p\rangle$ represent *both* the *type* of eigenket (respectively position and momentum in this case) and also the eigenvalue in the sense that $\hat{Q}|q\rangle = q|q\rangle$ and $\hat{P}|p\rangle = p|p\rangle$. As we shall see shortly, our choice of notation will allow us to write some equations in a more compact and elegant form. We shall designate the vector space generated by these basis as $\boldsymbol{W}^{(\infty)}$. This is clearly *not* a Hilbert space since $\boldsymbol{W}^{(\infty)}$ accommodates in it, "generalized vectors" as $|q(x)\rangle$ and $|p(x)\rangle$ with "infinite norm". A rigorous foundation for this construction can be given within the so called Rigged Vector Space formalism (see [10] for more details on this issue). Later on, we will present a natural construction of these spaces as a continuous heuristic limit of analogous well-defined finite dimensional spaces originally due to Schwinger. Let $\hat{V}_\xi$ and $\hat{U}_\eta$ $(-\infty < \xi, \eta < \infty)$ be a pair of unitary operators that implement the one parameter abelian group of translations on respectively the position and momentum basis in the sense that

$$\hat{V}_\xi|q(x)\rangle = |q(x-\xi)\rangle \quad \text{and} \quad \hat{U}_\eta|p(x)\rangle = |p(x+\eta)\rangle \tag{7}$$

with

$$\hat{V}_\xi \cdot \hat{V}_{\xi'} = \hat{V}_{\xi+\xi'}\,, \quad \hat{V}_\xi^\dagger = \hat{V}_{-\xi}\,, \quad \hat{U}_\eta \cdot \hat{U}_{\eta'} = \hat{U}_{\eta+\eta'} \quad \text{and} \quad \hat{U}_\eta^\dagger = \hat{U}_{-\eta} \tag{8}$$

The hermitian generators of $\hat{V}_\xi$ and $\hat{U}_\eta$ are respectively the momentum and position observables:

$$\hat{V}_\xi = e^{i\xi\hat{P}} \quad \text{and} \quad \hat{U}_\eta = e^{i\eta\hat{Q}} \tag{9}$$

The translation operators also obey the so called Weyl relation:

$$\hat{V}_\xi \cdot \hat{U}_\eta = e^{i\xi\eta}\hat{U}_\eta \cdot \hat{V}_\xi \tag{10}$$

which can be thought as an exponentiated version of the familiar Heisenberg relation:

$$[\hat{Q},\hat{P}] = i\hat{I} \tag{11}$$

The basis $|q(x)\rangle$ and $|p(x)\rangle$ are both *complete* and *normalized* in the sense that

$$\hat{I} = |q(x)\rangle\langle q(x)| = \int_{-\infty}^{+\infty} dx|p(x)\rangle\langle p(x)| \quad \text{and} \\ \langle q(x)|q(x')\rangle = \langle p(x)|p(x')\rangle = \delta(x-x') \tag{12}$$

The overlap between the position and momentum eigenstate is given by the well-known plane wave function:

$$\langle q(x)|p(x')\rangle = \frac{1}{\sqrt{2\pi}} e^{ixx'} \tag{13}$$

Any arbitrary abstract state $|\psi\rangle$ has a "position wave function" given by $\langle q(x)|\psi\rangle$ and a momentum basis function given by $\langle p(x)|\psi\rangle$. The relation between both is given by the Fourier transform operator which is a unitary operator that takes one basis to another as

$$\hat{F}|q(x)\rangle = |p(x)\rangle \tag{14}$$

This operator can be defined in a more elegant and compact manner as:

$$\hat{F} = \int_{-\infty}^{+\infty} dx|p(x)\rangle\langle q(x)| \tag{15}$$

Note that the above equation could not have been written in such way within the conventional notation. We also have that $\hat{F}^\dagger|p(x)\rangle = |q(x)\rangle$ and that $\hat{F}$ is unitary, which means that $\hat{F}^\dagger\hat{F} = \hat{I}$. The Fourier transform of $\langle q(x)|\psi\rangle$ is then

$$\langle q(x)|\hat{F}^\dagger|\psi\rangle = \langle p(x)|\psi\rangle \tag{16}$$

The effect of the inverse transform Fourier operator over the position basis is given by

$$\hat{F}^\dagger|q(x)\rangle = \int_{-\infty}^{+\infty} dx'|q(x')\rangle\langle p(x')|q(x)\rangle \tag{17}$$

By the form of the plane wave function, it is easy to see that $\langle p(x')|q(x)\rangle = \langle q(x')|p(-x)\rangle$ so that:

$$\hat{F}^\dagger|q(x)\rangle = \int_{-\infty}^{+\infty} dx'|q(x')\rangle\langle q(x')|p(-x)\rangle = |p(-x)\rangle \tag{18}$$

Analogously, we may write

$$\hat{F}^2|q(x)\rangle = \hat{F}|p(x)\rangle = \int_{-\infty}^{+\infty} dx'|p(x')\rangle\langle q(x')|p(x)\rangle = \\ = \int_{-\infty}^{+\infty} dx'|p(x')\rangle\langle p(x')|q(-x)\rangle \tag{19}$$

which means that

$$\hat{F}^2|q(x)\rangle = |q(-x)\rangle \tag{20}$$

This implies that $\hat{F}^2$ is nothing else but the *spatial inversion operator* in $\boldsymbol{W}^{(\infty)}$. Analogously one can deduce in a similar manner that

$$\hat{F}^2|p(x)\rangle = |p(-x)\rangle \tag{21}$$

This also implies that

$$\hat{F}^4 = \hat{I} \tag{22}$$

which shows that the spectrum of the Fourier operator $\hat{F}$ consists of the *fourth roots of unity*. We shall return to this issue in a later section. In a similar manner it is not difficult to show that

$$\begin{array}{ccc} \hat{F}^{\dagger}\cdot\hat{U}_{\xi}\cdot\hat{F}=\hat{V}_{\xi}^{\dagger} & \text{and} & \hat{F}^{\dagger}\cdot\hat{V}_{\xi}\cdot\hat{F}=\hat{U}_{\xi} \\ \hat{F}\cdot\hat{V}_{\xi}\cdot\hat{F}^{\dagger}=\hat{U}_{\xi}^{t} & \text{and} & \hat{F}\cdot\hat{U}_{\xi}\cdot\hat{F}^{\dagger}=\hat{V}_{\xi} \end{array} \tag{23}$$

which implies the following two very important relations

$$\hat{F}^{2}\cdot\hat{V}_{\xi}\cdot\hat{F}^{2}=\hat{V}_{\xi}^{\dagger} \quad \text{and} \quad \hat{F}^{2}\cdot\hat{U}_{\xi}\cdot\hat{F}^{2}=\hat{U}_{\xi}^{\dagger} \tag{24}$$

### *2.2 An intrinsic formulation for the Weyl-Wigner operator and Transform*

We shall designate the space of linear operators over $\boldsymbol{W}$ as $\boldsymbol{T}_{1}^{1}(\boldsymbol{W})$ and denote the "vectors", (the elements) $\hat{A},\hat{B},\hat{C},\ldots$ of $\boldsymbol{T}_{1}^{1}(\boldsymbol{W})$ respectively by $|\hat{A}\,),|\hat{B}\,),|\hat{C}\,),\ldots$ Following Schwinger [11], we introduce the well-known hermitian inner product (also known as the Hilbert-Schmidt inner product) in $\boldsymbol{T}_{1}^{1}(\boldsymbol{W})$ given by:

$$(\,\hat{A}\mid\hat{B}\,)=tr(\hat{A}^{\dagger}.\hat{B}) \tag{25}$$

The space $\boldsymbol{T}_{1}^{1}(\boldsymbol{W})$ has an additional algebraic structure which turns it into an operator algebra. In fact, we have the following product $|\,\hat{A}\,).|\,\hat{B}\,)=|\,\hat{A}.\hat{B}\,)$ where $\hat{A}.\hat{B}$ is the usual operator product between $\hat{A}$ and $\hat{B}$. In this way we may identify the elements of $\boldsymbol{T}_{1}^{1}(\boldsymbol{W})$ with the elements of $\boldsymbol{T}_{1}^{1}(\boldsymbol{T}_{1}^{1}(\boldsymbol{W}))$**,** in an unique manner through the obvious inclusion map

$$\begin{array}{c} \boldsymbol{i}:\boldsymbol{T}_{1}^{1}(\boldsymbol{W})\rightarrow\boldsymbol{T}_{1}^{1}\left(\boldsymbol{T}_{1}^{1}(\boldsymbol{W})\right) \\ \hat{A}\equiv|\hat{A}\,)\;\mapsto\;\overset{\frown}{A} \end{array} \tag{26}$$

Such that $\overset{\frown}{A}\,|\hat{B})=|\hat{A}.\hat{B})$. We will usually allow ourselves a slight abuse of language by dismissing any explicit mention to this inclusion map. For instance, consider the following identifications of the Identity Operator: $\hat{I}\equiv|\,\hat{I}\,)\overset{i}{\equiv}\overset{\frown}{\boldsymbol{I}}$ **.** In this way one may define a set of operators $\hat{X}(\alpha)\equiv|\hat{X}(\alpha)\,)$ to be *complete* in $\boldsymbol{T}_{1}^{1}(\boldsymbol{W})$ (where $\alpha$ is a variable that takes values in some appropriate "index set") if

$$\int d\alpha|\hat{X}(\alpha)\,)(\,\hat{X}(\alpha)\,|=\overset{\frown}{I}\equiv\hat{I}\equiv|\,\hat{I}\,) \tag{27}$$

Where $d\alpha$ is an appropriate *measure* in the index set. Orthonormality can written as

$$\big(\hat{X}(\alpha)\big|\hat{X}(\beta)\big)=\;tr(\hat{X}^{\dagger}(\alpha).\hat{X}(\beta))=\delta(\alpha-\beta) \tag{28}$$

As an example, consider $|q(x)\rangle\langle p(y)|\equiv|x,y)$ with $x,y\,\in\,\mathbb{R}^{2}$. This set of $\mathbb{R}^{2}$-valued operators is indeed orthonormal in the sense that $(x,y|x',y')=\delta(x-x')\delta(y-y')$. It is not difficult to prove that $\big(x,y\big|\hat{A}\big)=\big\langle q(x)\big|\hat{A}\big|p(y)\big\rangle$ and $\iint dxdy|x,y)(x,y|=\overset{\frown}{I}$. These results can be used to prove another very convenient completeness relation for a set $\{\hat{X}(\alpha)\}$ as

$$\int d\alpha\hat{X}^{\dagger}(\alpha).\hat{A}.\hat{X}(\alpha)=\big(tr\hat{A}\big)\hat{I}\quad\text{for}\quad\forall\,\hat{A}\,\in\,\boldsymbol{T}_{1}^{1}(\boldsymbol{W}) \tag{29}$$

We can now define a set of operators (the so called Weyl-Wigner operators) also parameterized by the phase space plane $(x,y)\in\,\mathbb{R}^{2}$ in the following intrinsic manner

$$\hat{\Delta}(x,y)=2\hat{V}_{y}^{\dagger}.\hat{U}_{2x}.\hat{V}_{y}^{\dagger}.\hat{F}^{2} \tag{30}$$

With aid of equations (23), one can easily prove the following important properties:

$$\hat{\Delta}^{\dagger}(x,y)=\hat{\Delta}(x,y) \qquad \text{(hermiticity)}$$

$$tr(\hat{\Delta}(x,y)) = (\hat{\Delta}(x,y)|\,\hat{I}) = 1$$ (Unit trace)

$$(\hat{\Delta}(x,y)|\,\hat{\Delta}(x',y')) = 2\pi\delta(x-x')\delta(y-y')$$ (Orthonormality of $\{{}^{1}/_{\sqrt{2\pi}}\hat{\Delta}(x,y)\}$)

$$\frac{1}{2\pi}\iint dxdy|\,\hat{\Delta}(x,y))(\hat{\Delta}(x,y)| = \hat{I}$$ (1st form for the Completeness of $\{{}^{1}/_{\sqrt{2\pi}}\hat{\Delta}(x,y)\}$)

$$\frac{1}{2\pi}\iint dxdy\hat{\Delta}(x,y).\hat{A}.\hat{\Delta}(x,y) = (tr\hat{A})\hat{I}, \text{for all} \hat{A}$$ (2nd form of the Completeness of $\{{}^{1}/_{\sqrt{2\pi}}\hat{\Delta}(x,y)\}$)

Some other properties (not so well-known) for the Weyl-Wigner operators are the following:

$$\hat{\Delta}^2(x,y) = 4\hat{I}, \qquad \hat{F}^2.\hat{\Delta}(x,y).\hat{F}^2 = \hat{\Delta}(-x,-y) \qquad \text{and} \qquad \hat{\Delta}(0,0) = 2\hat{F}^2$$

At this point, one can define the Weyl-Wigner Transform of an arbitrary Operator as

$$W\{\hat{A}\}(x,y) = (\hat{\Delta}(x,y)|\hat{A}) = a(x,y) \tag{31}$$

The transform of $\hat{A}$ is in general a *complex function* $a(x,y)$ of the phase plane, but if $\hat{A}$ is hermitian, then $a(x,y)$ is clearly *real-valued*. One is tempted to see this transform as a map between quantum observables to “classical observables” in some sense. In fact, as we shall soon see, there is a certain sense where in the $\hbar \to 0$ limit, one can see that $a(x,y)$ goes indeed to the expected classical observable. Wigner introduced this transform to map density operators of mixed states to classical probability densities over phase space, but the fact is that these densities obey all axioms for a true probability distribution on phase space with the *exception* of *positivity*. In this way, the *negativity* of the transform of a density operator signals for a *departure of classicality* of the mixed state or some kind of measure of “quanticity” of the state.

### 2.3 – The Classical Limit

The inner product between two operators $\hat{A}$ and $\hat{B}$ can be written in terms of their WW transforms as

$$(\hat{A}|\hat{B}) = \frac{1}{2\pi}\int dxdy\bar{a}(x,y)b(x,y) \tag{32}$$

Where $\bar{a}(x,y)$ is the c.c. of $a(x,y)$. The transform of a product of two arbitrary operators can be seen (after a tedious calculation) as

$$(\hat{\Delta}(x,y)|\hat{A}\hat{B}) = a(x,y)\exp\left[-\frac{\mathrm{i}}{2}\left(\frac{\overleftarrow{\partial}}{\partial x}\frac{\overrightarrow{\partial}}{\partial y} - \frac{\overleftarrow{\partial}}{\partial y}\frac{\overrightarrow{\partial}}{\partial x}\right)\right]b(x,y) \tag{33}$$

Where the “arrows” above the partial derivative operators indicate “which” function it operates on. With this result, it is not difficult to guess the very suggestive form of the WW transform of the *commutator* of two operators and (we momentarily restore $\hbar$ here explicitly) the following fact that the “classical limit” can be understood as

$$\begin{aligned}\lim_{\hbar\to 0}(\hat{\Delta}(x,y)|[\hat{A}\hat{B}]) &= \lim_{\hbar\to 0}\frac{i}{\hbar}a(x,y)\sin\left[-\frac{\mathrm{i}}{2}\left(\frac{\overleftarrow{\partial}}{\partial x}\frac{\overrightarrow{\partial}}{\partial y} - \frac{\overleftarrow{\partial}}{\partial y}\frac{\overrightarrow{\partial}}{\partial x}\right)\right]b(x,y) \\ &= -\{a(x,y),b(x,y)\}\end{aligned} \tag{34}$$

One recognizes the last term of the above equation as the Poisson bracket of the classical observables $a(x,y)$ and $b(x.y)$. This operation defines a *non-commutative* algebra over the real functions defined on phase space. It is also known as the *star-product* and it forms the basis of the non-commutative geometric point of view for quantization.

### *2.4 – The issue of the number of degrees of freedom*

The extension of the *continuous* WW formalism to two or more degrees of freedom is straightforward: For instance, let $|q(x)\rangle$ and $|q(y)\rangle$ be complete position eigenbasis respectively for the $x$ and $y$ directions, then a point $\vec{r}$ of the plane is clearly represented by the tensor product state $|q(\vec{r})\rangle = |q(x)\rangle \otimes |q(y)\rangle$. The 2D translation operator

$$\hat{V}_{\vec{\xi}} = \exp(i\vec{P}.\vec{\xi}) = \exp(i\hat{P}_x\,\xi_x) \otimes \exp(i\hat{P}_y\,\xi_y) \tag{35}$$

acts upon the $|q(\vec{r})\rangle$ basis by the expected manner as $\hat{V}_{\vec{\xi}}|q(\vec{r})\rangle = |q(\vec{r}-\vec{\xi})\rangle$. This can be clearly carried out for any number of degrees of freedom in the same way. For finite dimensional quantum spaces, things are *not* as simple as we shall see further ahead.

# 3-The Geometry of Quantum Mechanics

(The main references for this chapter are [6], [12] and [13])

*The Projective space CP(N)*

Let us consider the $(N+1)$-dimensional quantum space $\boldsymbol{W}^{(N+1)}$. This a complex vector space together with an *anti-linear map* between $\boldsymbol{W}^{(N+1)}$ and its dual $\overline{\boldsymbol{W}}^{(N+1)}$:

$$\begin{aligned} \langle\ \ \rangle :\ \boldsymbol{W}^{(N+1)} &\rightarrow \overline{\boldsymbol{W}}^{(N+1)} \\ |\psi\rangle\ &\mapsto\ \langle\psi| = (|\psi\rangle)^{\dagger} \end{aligned} \tag{36}$$

that takes each *ket* state to its associated *bra* state through the familiar "dagger" operation. The inner product between $|\psi_1\rangle$ and $|\psi_2\rangle$ can then be defined as the natural action of $\langle\psi_1|$ over $|\psi_2\rangle$ in the usual Dirac notation$\langle\psi_1|\psi_2\rangle$. Following [6], one can define *Euclidean* and *symplectic* metric structures on $\boldsymbol{W}^{(N+1)}$ through the relations

$$\boldsymbol{G}(|\psi_1\rangle, |\psi_2\rangle) = Re(\langle\psi_1|\psi_2\rangle) \quad \text{and} \quad \boldsymbol{\Omega}(|\psi_1\rangle, |\psi_2\rangle) = Im(\langle\psi_1|\psi_2\rangle) \tag{37}$$

From fundamental postulates of quantum mechanics, it is clear that a state vector $|\psi\rangle \in \boldsymbol{W}^{(N+1)}$ is *physically indistinguishable* from the state vector $\lambda|\psi\rangle$ (for arbitrary $\lambda \in \mathbb{C}$. Thus, the *true* physical space of states of the theory is the so called "space of rays" or the complex projective space $\mathbb{C}P(N)$ defined by the quotient of $\boldsymbol{W}^{(N+1)}$ by the above physically motivated equivalence relation.

Given an orthonormal basis $\{|u_\sigma\rangle\}$ $\sigma = 0,1,2,\ldots$, then a general state vector can be expanded as $|\psi\rangle = |u_\sigma\rangle z^\sigma$. One can map this state to a sphere $S^{(2N+1)}$ with radius given by $\overline{z}_\sigma z^\sigma = r^2$. We introduce projective coordinates $\xi^i = z^i/z^0$ on $\mathbb{C}P(N)$ so that $z^0 = re^{i\varphi}/\left(1+\bar{\xi}_i\xi^i\right)^{1/2}$, with $i = 1,\ldots,n$, where $\varphi$ is an arbitrary phase factor. The Euclidean metric in $\boldsymbol{W}^{(N+1)}$ (seen here as a $(2n+2)$-dimensional *real* vector space) can be written as $dl^2\left(\boldsymbol{W}^{(N+1)}\right) = d\bar{z}_\sigma dz^\sigma = dr^2 + r^2 dS^2(S^{(2N+1)})$, where

$$dS^2\left(S^{(2N+1)}\right) = (d\varphi - A)^2 + ds^2(\mathbb{C}P(N)) \tag{38}$$

is the squared distance element over the space of normalized vectorsandthe one-form $A = \frac{i}{2}\frac{\bar{\xi}_i d\xi^i - d\bar{\xi}_i \xi^i}{(1+\bar{\xi}_i\xi^i)}$ is the well known abelian connection of the $U(1)$ bundle over $\mathbb{C}P(N)$[12]**.**The metric $ds^2(\mathbb{C}P(N))$ over the space of rays in projective coordinates is given explicitly by:

$$ds^2(\mathbb{C}P(N)) = \frac{i}{2}\left[\frac{(1+\bar{\xi}_i\xi^i)\delta_j^k - \bar{\xi}_k\xi^j}{\left(1+\bar{\xi}_i\xi^i\right)^2}\right] d\bar{\xi}_j \xi^k \tag{39}$$

an d**Ω**= $dA$ is a 2-form defined over $\mathbb{C}P(N)$ which makes it a symplectic manifold. These quantum symplectic spaces are physically very different in origin from their Newtonian counterparts. The Newtonian phase spaces are cotangent bundles over some configuration manifold and as such they are always non-compact manifolds, while projective spaces are *compact* for finite $N$. For instance, $\mathbb{C}P(1)$ can be identified with a 2-sphere. The quantum projective spaces also have an additional structure given by their Riemannian metric, which is absent in the Newtonian case. Both these structures are compatible in a precise sense that makes these complex projective spaces examples of what is called a *Kahler* manifold. A more natural and intuitive pictorial representation of these structures can be seen easily in the figure below

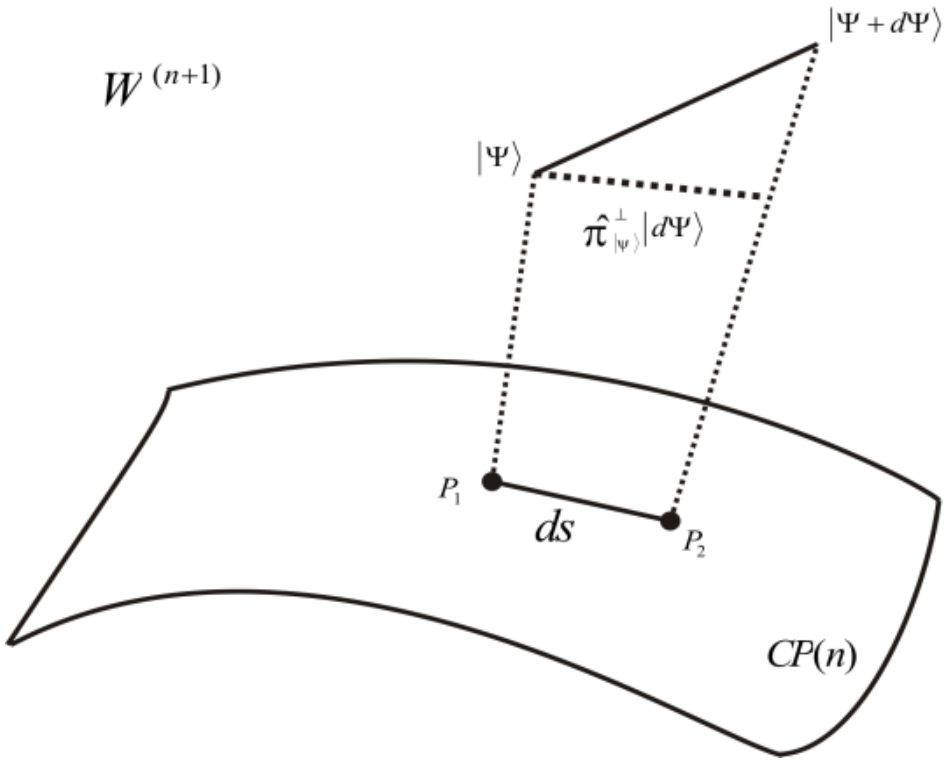

**Figure 1 - Pictorial representation of the quantum space of states**

The points $P_1$ and $P_2 \in \mathbb{C}P(N)$ are the projections respectively from two infinitesimally nearby normalized state vectors $|\psi\rangle$ and $|\psi + d\psi\rangle$. It is natural to define then, the squared distance between $P_1$ and $P_1$ as the projection of $|d\psi\rangle$ in the "orthogonal direction" of $|\psi\rangle$, that is, the projection given by the projective operator $\hat{\pi}_{|\psi\rangle} = \hat{I} - |\psi\rangle\langle\psi|$. It is then easy to see that

$$ds^2\big(\mathbb{C}P(N)\big) = \langle d\psi|d\psi\rangle - \langle d\psi|\psi\rangle\langle\psi|d\psi\rangle \quad (40)$$

which is an elegant and coordinate independent manner to express the metric over $\mathbb{C}P(N)$.The time-evolution of a physical state in $\mathbb{C}P(N)$ is given by the projection of the linear Schrödinger evolution in $\boldsymbol{W}^{(N+1)}$ which results in a Hamiltonian (classical-like) evolution in the space of rays.

# 4 - Coherent States and the Quantum Symplectic Group

(The main references for this chapter are[14], [15] and [16])

## *4.1 The spectrum of the Fourier Transform operator*

Let $\hat{a}$ and $\hat{a}^\dagger$ be respectively the usual annihilation and creation operators defined as [9]

$$\hat{a} = \frac{1}{\sqrt{2}}(\hat{Q} + i\hat{P}) \quad \text{and} \quad \hat{a}^\dagger = \frac{1}{\sqrt{2}}(\hat{Q} - i\hat{P}) \quad (41)$$

It follows immediately from the Heisenberg commutation relations that $[\hat{a}, \hat{a}^\dagger] = \hat{I}$, and if we define also the hermitian number operator $\hat{N} = \hat{a}^\dagger\hat{a}$ it is not difficult to derive the well-known relations:

$$[\hat{N}, \hat{a}] = -\hat{a} \quad \text{and} \quad [\hat{N}, \hat{a}^\dagger] = \hat{a}^\dagger \quad (42)$$

which implies the also well-known spectrum $\hat{N}|n\rangle = n|n\rangle$, with $n = 0,1,2, \ldots$ of the number operator and the “up and down-the-ladder” action for $\hat{a}^\dagger$ and $\hat{a}$:

$$\hat{a}|n\rangle = \sqrt{n}|n-1\rangle \quad \text{and} \quad \hat{a}^\dagger|n\rangle = \sqrt{n+1}|n+1\rangle \tag{43}$$

and also where the ground-state or vacuum state $|0\rangle$ is *annihilated* by $\hat{a}$:

$$\hat{a}|0\rangle = 0$$

The above algebraic equation for the vacuum state can be written in the position and momentum basis as the following differential equations:

$$\langle q(x)|\hat{a}|0\rangle = \frac{1}{\sqrt{2}}\left(x + \frac{d}{dx}\right)\langle q(x)|0\rangle = \langle p(x)|\hat{a}|0\rangle = \frac{i}{\sqrt{2}}\left(x + \frac{d}{dx}\right)\langle p(x)|0\rangle = 0$$

Which gives us the normalized Gaussian functions $\langle q(x)|0\rangle = \langle p(x)|0\rangle = (\pi)^{-1/4}e^{-x^2/2}$.

From section 2, we know that $\hat{F}|q(x)\rangle = |p(x)\rangle$ so we conclude that the vacuum state is a fixed point of the Fourier Transform operator $\hat{F}|0\rangle = |0\rangle$. The infinitesimal versions of equations (23) are simply

$$\hat{F}^\dagger\hat{a}\hat{F} = i\hat{a} \quad \text{and} \quad \hat{F}^\dagger\hat{a}^\dagger\hat{F} = -i\hat{a}^\dagger \tag{44}$$

Which implies that $[\hat{F}, \hat{N}] = 0$. Thus, the Fourier transform operator commutes with the number operator so they share the same eigenstates. From the fixed point condition of the vacuum state together with the equations above, it is not difficult to see that

$$\hat{F}|n\rangle = (i)^n|n\rangle \quad \text{and} \quad \hat{F} = e^{i(\pi/2)\hat{N}} \tag{45}$$

Which implies that $\hat{N}$ is the hermitian generator of the Fourier transform. In fact, at this point, it is a quite obvious move to introduce a complete *rotation operator* in phase space known as the *Fractional Fourier Operator* given by $\hat{F}_\theta = e^{i\theta\hat{N}}$[4].The Fourier transform operator is then recognized as a special case for $\theta = \pi/2$.

### 4.2 The Quantum Linear Symplectic Transforms

It is also natural to extend the Fractional Fourier transform to a complete set of quantum *symplectic transforms*, those unitary transformation in $\boldsymbol{W}^{(\infty)}$ that implement a representation of the group of area preserving linear maps of the classical phase plane. This is the non-abelian $SL(2,\mathbb{R})$ group. Probably the best way to visualize this is through the identification of the phase plane with the complex plane via the standard complex-valued coherent states defined by the following change of variables: $z = (1/\sqrt{2})(q + ip)$ and by defining the coherent states as $|z\rangle = \hat{D}[z]|0\rangle$, with the displacement $\hat{D}[z]$ operator given by

$$\hat{D}[z] = \frac{1}{2}\hat{\Delta}(z/2)\hat{F}^2 = e^{z\hat{a}^\dagger - \bar{z}\hat{a}} \tag{46}$$

The well-known "over-completeness" of the coherent state representation follows from the completeness of the $\hat{\Delta}(z)$ basis. The overlap between distinct coherent states is given by

$$\langle z|z'\rangle = \langle p,q|p',q'\rangle = e^{-\frac{1}{4}[(p-p')^2+(q-q')^2]}e^{i(p'q-pq')/2}$$

Note above, the *symplectic phase* proportional to the area in the phase plane defined by the vectors $(p,q)$ and $(p',q')$. It is also not difficult to show that indeed $\hat{F}_\theta|z\rangle = e^{i\theta\hat{N}}|z\rangle = |e^{i\theta}z\rangle$ which is a direct manner to represent the Fractional Fourier transform for arbitrary $\theta$. Since the generator of rotations is *quadratic* in the canonical observables $\hat{Q}$ and $\hat{P}$, one may try to write down all possible quadratic operators in these variables: $\hat{Q}^2, \hat{P}^2, \hat{Q}\hat{P}$ and $\hat{P}\hat{Q}$, but the last two are obviously non-Hermitian so we could change them to the following (Hermitian) linear combinations: $\hat{Q}\hat{P} + \hat{P}\hat{Q}$ and $i(\hat{Q}\hat{P} - \hat{P}\hat{Q})$. The last one is proportional to the identity operator because of the Heisenberg commutation relation, so this leaves us with *three* linear independent operators that we choose as

$$\hat{H}_0 = \frac{1}{2}(\hat{Q}^2 + \hat{P}^2) = \hat{N} + \frac{1}{2}\hat{I} = \hat{a}^\dagger\hat{a} + \frac{1}{2}\hat{I} \tag{47}$$

$$\hat{g} = \frac{1}{2}\{\hat{Q}, \hat{P}\} = \frac{1}{2}(\hat{Q}\hat{P} + \hat{P}\hat{Q}) = \frac{i}{2}[(\hat{a}^{\dagger})^2 - \hat{a}^2] \tag{48}$$

$$\hat{k} = \frac{1}{2}(\hat{Q}^2 - \hat{P}^2) = \frac{1}{2}[(\hat{a}^{\dagger})^2 + \hat{a}^2] \tag{49}$$

These three generators implement in $\boldsymbol{W}^{(\infty)}$, the algebra $sl(2,\mathbb{R})$ of $SL(2,\mathbb{R})$. The $\hat{g}$ operator is nothing but the *squeezing* generator from quantum optics [17]. Indeed, the scale operator

$\hat{S}_{\xi} = e^{i ln\xi \hat{g}}$ generated by $\hat{g}$ act upon the position and momentum basis respectively as $\hat{S}_{\xi}|q(x)\rangle = \sqrt{\xi}|q(\xi x)\rangle$ and $\hat{S}_{\xi}|p(x)\rangle = \frac{1}{\sqrt{\xi}}|p(\frac{x}{\xi})\rangle$. The $\hat{k}$ operator generates *hyperbolic rotations*, that is, linear transformations of the plane that preserve an indefinite metric. It takes the hyperbola $x^2 - y^2 = 1$ into itself in an analogous way that the Euclidean rotation takes the circle $x^2 + y^2 = 1$ into itself. $SL(2,\mathbb{R})$ is the Lie Group of all area preserving linear transformations of the plane, so we can identify it with the $2 \times 2$ real matrices with unit determinant. Since $\det(e^X) = e^{tr(X)}$, we can also identify the algebra $sl(2,\mathbb{R})$ with all $2 \times 2$ real matrices with null trace. Thus, it is natural to make the following choice for a basis in this algebra:

$$\hat{X}_1 = \hat{\sigma}_1, \quad \hat{X}_2 = i\hat{\sigma}_2 \quad \text{and} \quad \hat{X}_3 = \hat{\sigma}_3$$

where we have written (for practical purposes) the elements of the algebra in terms of the well-known Pauli matrices. This is very adequate because physicists are familiar with the commutation relations of the Pauli matrices because they form a two-dimensional representation of the angular momentum algebra and we can make use of these algebraic relations to completely characterize the $sl(2,\mathbb{R})$ algebra. In fact, the mapping described by the table below relates these algebra elements directly to the algebra of their representation carried on $\boldsymbol{W}^{(\infty)}$:

| **generators of $sl(2,\mathbb{R})$** | **generators of the representation** |
|---|---|
| $\hat{X}_1 = \hat{\sigma}_1$ | $-i\hat{k}$ |
| $\hat{X}_2 = i\hat{\sigma}_2$ | $-i\hat{H}_0$ |
| $\hat{X}_3 = \hat{\sigma}_3$ | $-i\hat{g}$ |

With a bit of work, it is not difficult to convince oneself that these mapped elements indeed obey identical commutation relations.

## 5 – Applications: Weak Values and Quantum Mechanics in Phase Space

The weak value of a quantum system was introduced by Aharonov, Albert and Vaidman (A.A.V.) in [18] based on the two-state formalism for quantum mechanics [19,20] and it generalizes the concept of an expectation value for a given observable. Let the initial state of a product space $\boldsymbol{W} = \boldsymbol{W}^{(S)} \otimes \boldsymbol{W}^{(M)}$ be given the product state $|\Psi\rangle = |\alpha\rangle \otimes |\phi^{(i)}\rangle$ where $|\alpha\rangle$ is the *pre-selected* state of the system and $|\phi^{(i)}\rangle$ is the *initial state* of the apparatus. Suppose further that a "weak Hamiltonian" governs the interaction between the system and the measuring apparatus as:

$$\hat{H}_{int} = \epsilon\delta(t - t_0)\hat{O} \otimes \hat{P}(\epsilon \to 0) \tag{50}$$

where $\hat{O}$ is an arbitrary observable to be measured in the system. After the ideal instantaneous interaction that models this von-Neumann (weak) measurement [21], suppose we *post-select* a certain final state $|\beta\rangle$ of the system after performing a *strong* measurement on it. In this case, the final state is clearly given by

$$|\phi^{(f)}\rangle = (\langle\beta| \otimes \hat{I})e^{-i\epsilon\hat{O}\otimes\hat{P}}(|\alpha\rangle \otimes |\phi^{(i)}\rangle) \approx \langle\beta|\alpha\rangle(1 - i\epsilon O_w)|\phi^{(i)}\rangle \tag{51}$$

where

$$O_w = \frac{\langle \beta | \hat{O} | \alpha \rangle}{\langle \beta | \alpha \rangle}$$

is the weak value of the observable $\hat{O}$ for these particular chosen pre and post-selected states. Note that the weak value $O_w$ of the observable is, in general, an arbitrary complex number. Note also that, though $|\phi^{(i)}\rangle$ is a *normalized* state, the $|\phi^{(f)}\rangle$ state vector in general, is *not* normalized. In the original formulation of (A.A.V.), the momentum $\hat{P}$ acts upon the measuring system, implementing a small translation of the initial wave function in the position basis, but which can be measured from the mean value of the results of a large series of identical experiments. That is, the expectation value of the position operator $\hat{Q}$ over a large ensemble with the same pre and post selected states. One can generalize this procedure with [22] by taking an arbitrary operator $\hat{M}$ in the place of $\hat{Q}$ as the observable of $\boldsymbol{W}^{(M)}$ to be measured. In this case, the usual expectation values of $\hat{M}$ for the initial and final states $|\phi^{(i)}\rangle$ and $|\phi^{(f)}\rangle$ are respectively:

$$\langle \hat{M} \rangle_{(i)} = \langle \phi^{(i)} | \hat{M} | \phi^{(i)} \rangle \quad \text{and} \quad \langle \hat{M} \rangle_{(f)} = \frac{\langle \phi^{(f)} | \hat{M} | \phi^{(f)} \rangle}{\langle \phi^{(f)} | \phi^{(f)} \rangle} \tag{52}$$

and the difference between these expectation values (the shift of $\hat{M}$) to first order in ε is given in general by [22]:

$$\begin{aligned} \Delta \hat{M} &= \langle \hat{M} \rangle_{(f)} - \langle \hat{M} \rangle_{(i)} \\ &= \epsilon[\big(Im(O_w)\big)\big(\langle \phi^{(i)} | \{\hat{M}, \hat{P}\} | \phi^{(i)} \rangle - 2\langle \phi^{(i)} | \hat{P} | \phi^{(i)} \rangle \langle \phi^{(i)} | \hat{M} | \phi^{(i)} \rangle\big) \\ &\quad - i\big(Re(O_w)\big)\langle \phi^{(i)} | [\hat{M}, \hat{P}] | \phi^{(i)} \rangle] \end{aligned}$$

For the choice $\hat{M} = \hat{Q}$ and also by using the Heisenberg picture for the time evolution and choosing the most general Hamiltonian $\hat{H} = \frac{1}{2m}\hat{P}^2 + V(\hat{Q})$ for the measuring system one can derive the following shift (using the $sl(2, \mathbb{R})$ algebra and the Heisenberg commutation relation):

$$\Delta \hat{Q} = \epsilon[Re(O_w) + Im(O_w) m \frac{d}{dt}\left(\delta^2_{|\phi^{(i)}\rangle} \hat{Q}\right)] \tag{53}$$

Where $\delta^2_{|\phi^{(i)}\rangle} \hat{Q}$ is the uncertainty of the initial state $|\phi^{(i)}\rangle$ and analogously for $\hat{M} = \hat{P}$, one arrives at

$$\Delta \hat{P} = 2\epsilon Im(O_w)\left(\delta^2_{|\phi^{(i)}\rangle} \hat{P}\right) \tag{54}$$

This result is clearly asymmetric because of the choice of the translation generator $\hat{P}$ in the interaction Hamiltonian. Note also that from the above equations one can see that it is impossible to extract the real and imaginary values of the weak value with the measurement of $\Delta \hat{Q}$ only, because both of these numbers are absorbed in a same real number. It is necessary to measure $\Delta \hat{P}$ (besides knowing the values of $\frac{d}{dt}\left(\delta^2_{|\phi^{(i)}\rangle} \hat{Q}\right)$ and $\left(\delta^2_{|\phi^{(i)}\rangle} \hat{P}\right)$. There is no reason why one should need to choose $\hat{P}$ or $\hat{Q}$ in the weak measurement Hamiltonian. One may choose any of the symplectic generators making use of the full symmetry of the $SL(2, \mathbb{R})$ group. The $\hat{P}$ and $\hat{Q}$ operators generate translations in phase space, but one can implement any area preserving transformation in the plane by also using observables that are quadratic in the momentum and position observables. By making use of the freedom of choice of an arbitrary initial state vector $|\phi^{(i)}\rangle$ one can choose also an interaction Hamiltonian of the following form:

$$\hat{H}_{int} = \epsilon \delta(t - t_0) \hat{O} \otimes \hat{R} \qquad (\epsilon \to 0) \tag{55}$$

where $\hat{R}$ is any element of the algebra $sl(2, \mathbb{R})$, so it is the generator of an arbitrary symplectic transform of the measuring system. In this way the generalized $\Delta \hat{M}$ shift in these conditions is given by:

$$\Delta\hat{M} = \epsilon[\left(Im(O_w)\right)\left(\langle\phi^{(i)}|\{\hat{M},\hat{R}\}|\phi^{(i)}\rangle - 2\langle\phi^{(i)}|\hat{R}|\phi^{(i)}\rangle\langle\phi^{(i)}|\hat{M}|\phi^{(i)}\rangle\right) - i\left(Re(O_w)\right)\langle\phi^{(i)}|[\hat{M},\hat{R}]|\phi^{(i)}\rangle]$$

By making the choice $\hat{M} = \hat{R}$, one arrives at:

$$\Delta\hat{R} = 2\epsilon Im(O_w)\left(\delta^2_{|\phi^{(i)}\rangle}\hat{R}\right) \tag{56}$$

For the second observable, we could choose any observable that does *not* commute with $\hat{R}$. This is because the main idea is to choose a "conjugate" variable to $\hat{R}$ in a similar way that occurs with the $(\hat{Q},\hat{P})$ pair. So one obvious choice is to pick the number operator $\hat{N}$ in the place of $\hat{R}$. Since $\hat{N}$ is the generator of Euclidean rotations in phase space, the annihilator operator $\hat{a}$ seems a natural candidate operator (though not hermitian) to go along with $\hat{N}$. With this choice of $\hat{M} = \hat{a}$ it is not difficult to calculate the shift for the annihilator operator:

$$\Delta\hat{a} = \epsilon[-iO_w\langle\phi^{(i)}|\hat{a}|\phi^{(i)}\rangle + 2Im(O_w)\left(\langle\phi^{(i)}|\hat{N}\hat{a}|\phi^{(i)}\rangle - \langle\phi^{(i)}|\hat{N}|\phi^{(i)}\rangle\langle\phi^{(i)}|\hat{a}|\phi^{(i)}\rangle\right)]$$

In most models of weak measurements, the initial state of the measuring system is chosen to be a Gaussian state and the weak interaction promotes a small translation of its peak. In realistic quantum optical implementations of the measuring system, it is reasonable to choose the initial state of the system as a coherent state $|\phi^{(i)}\rangle = |z\rangle$. In this case, there is a dramatic simplification for the shift:

$$\Delta\hat{a} = i\epsilon O_w = \epsilon|O_w|e^{i(\theta_z+\theta_w-\pi/2)} \tag{57}$$

where $z = |z|e^{i\theta_z}$ and $O_w = |O_w|e^{i\theta_w}$. If we make the following convenient choice for the phase $\theta_z = \pi/2$ and rewrite the above equation in terms of the canonical pair $(\hat{Q},\hat{P})$, we arrive at a symmetric pair of equations for $\Delta\hat{Q}$ and $\Delta\hat{P}$:

$$\Delta\hat{Q} = \epsilon\sqrt{2}|z|Re(O_w) \quad \text{and} \quad \Delta\hat{P} = \epsilon\sqrt{2}|z|Im(O_w) \tag{58}$$

These equations do not depend on the quadratic dispersion or the time derivative of the quadratic dispersion of any observable and, in principle, one may "tune" the size of the $\epsilon|z|$ term despite how small $\epsilon$ may be by making $|z|$ large enough. This is of great practical importance for optical implementations of weak value measurements since $|z|$ for a quantized mode of an electromagnetic field is nothing else but the mean photon number in this mode for the coherent state $|z\rangle$[17].

## 6 - Applications: von-Neumann's pre-measurement, weak values and the geometry of quantum mechanics

In the last section we discussed von Neumann's model for a pre-measurement in the weak measurement limit in order to obtain a deeper understanding of the concept of a weak value in terms of a quantum phase space analysis of the *measuring apparatus* system. In this section we implement, in a certain sense, the *opposite* approach: We shall discuss certain geometric structures of the *measured* system based on previous work of Tamate et al [23]. Let $\boldsymbol{W} = \boldsymbol{W}^{(S)}\otimes\boldsymbol{W}^{(M)}$ be the state space formed by composing the subsystem $\boldsymbol{W}^{(S)}$ with the measuring subsystem $\boldsymbol{W}^{(M)}$. We will initially assume that the measured system is a discrete quantum variable of $\boldsymbol{W}^{(S)}$ defined by an observable $\hat{O} = |o_k\rangle o_k\langle o^k|$ (we use henceforth the sum convention).The measuring subsystem will be considered as a structure-less (no spin or internal variables) quantum mechanical particle in one dimension (further ahead we will also consider discrete measuring systems).Suppose the initial state of the total system be given by the following unentangled product state: $|\Psi^{(i)}\rangle = |\alpha\rangle\otimes|\phi^{(i)}\rangle$. After performing an ideal von-Neumann measurement through the interaction Hamiltonian $\hat{H}_{int} = \lambda\delta(t-t_0)\hat{O}\otimes\hat{P}$, the final state will be

$$(\hat{I}\otimes\langle q(x)|)|\Psi^{(f)}\rangle = |o_j\rangle\alpha^j\phi^{(i)}(x-\lambda o_j) \tag{59}$$

where $\phi^{(i)}(x) = \langle q(x)|\phi^{(i)}\rangle$ is the wave-function in the position basis of the measuring system. Note that a correlation in the final state of the total system is then established between the variable to be measured $o_j$with the continuous position variable of the measuring particle.This step of the von Neumann measurement prescription is called the *pre-measurement* of the system.

Consider now the measuring system as a *finite dimensional* quantum system $\boldsymbol{W}^{(M)}$. In particular, if $n = 2$, our measuring apparatus consists of a *single* qubit so that we can treat this two-level measuring system making explicit use of the $\mathbb{C}P(1)$ (Bloch sphere) geometry. A single qubit can be written in the Bloch sphere standard form as $|\theta,\varphi\rangle = \cos(\theta/2)|u_0\rangle + e^{i\varphi}\sin(\theta/2)|u_1\rangle$. The single *projective coordinate* in this case is $\xi = tan(\theta/2)e^{i\varphi}$ and, remarkably, we shall see that this complex number can actually be *measured* physically as a certain appropriate *weak value* for two level systems. Suppose now that the interaction happens in an arbitrary finite dimensional measuring system: $\boldsymbol{W} = \boldsymbol{W}^{(S)} \otimes \boldsymbol{W}^{(M)}$, that is $\dim\big(\boldsymbol{W}^{(M)}\big) = m$. The initial separable pure-state is $|\Psi^{(i)}\rangle = |\alpha\rangle \otimes |\phi^{(i)}\rangle$ and the finite momentum basis is given by $\{|v_\sigma\rangle\}, (\sigma = 0,1,\dots,m-1)$ so that the momentum observable can be expressed as $\hat{P} = |v_\sigma\rangle p_\sigma \langle v^\sigma|$. Again we model our instantaneous interaction with the interaction Hamiltonian $\hat{H}_{int} = \lambda\delta(t - t_0)\hat{O} \otimes \hat{P}$ so that our final entangled state is given by

$$|\Psi^{(f)}\rangle = |A_\sigma\rangle \otimes |v_\sigma\rangle \phi^\sigma \tag{60}$$

where $|A_\sigma\rangle = e^{-i\lambda p_\sigma \hat{O}}|\alpha\rangle$ and $|\phi^{(i)}\rangle = |v_\sigma\rangle\phi^\sigma$.The above entangled state clearly establishes a *finite* index correlation between $|A_\sigma\rangle \in \boldsymbol{W}^{(S)}$ and the finite momentum basis $|v_\sigma\rangle$. The total system is in the pure state $|\Psi^{(f)}\rangle\langle\Psi^{(f)}|$ and by *tracing out* the first subsystem, the measuring system will be:

$$\hat{\rho}_{|\Psi^{(f)}\rangle} = |v_\sigma\rangle \phi^\sigma \langle A^\tau|A_\sigma\rangle \bar{\phi}_\tau \langle A^\tau| \tag{61}$$

For a single qubit, one has$|\phi^{(i)}\rangle = \cos(\theta/2)|v_0\rangle + e^{i\varphi}\sin(\theta/2)|v_1\rangle$ with $\langle A^0|A_1\rangle = |\langle A^0|A_1\rangle| e^{-i\eta}$ so that we can compute the probability $P(\eta)$ of finding the second subsystem in a certain reference state $|\theta = \pi/2, \varphi = 0\rangle$ as

$$P(\eta) = tr(\hat{\rho}_{|\Psi^{(f)}\rangle}|\pi/2,0\rangle\langle\pi/2,0|) = \frac{1}{2} + \frac{1}{4}\langle A^0|A_1\rangle sin(\theta) cos(\varphi - \eta) \tag{62}$$

For a fixed $\theta$, this probability is *maximized* when $\varphi = \eta$. This fact can be used to measure the so called geometric phase $\eta = arg(\langle A^1|A_0\rangle)$ between the two indexed states $|A_0\rangle$ and $|A_1\rangle \in \boldsymbol{W}^{(S)}$. This definition of a geometric phase was originally proposed in 1956 by Pancharatnam [24] for optical states and rediscovered by Berry in 1984 [25] in his study of the adiabatic cyclic evolution of quantum states. In 1987, Anandan and Aharonov [26] gave a description of this phase in terms of geometric structures of the $U(1)$ fiber-bundle structure over the space of rays and of the symplectic and Riemannian structures in the projective space $\mathbb{C}P(N)$ inherited from the hermitian structure of $\boldsymbol{W}^{(S)}$.

Given the final state $|\Psi^{(f)}\rangle$, one may then "post-select" a chosen state $|\beta\rangle$ of $\boldsymbol{W}^{(S)}$. The resulting state is then clearly

$$|\Psi^{(f)}\rangle = C(|\beta\rangle\langle\beta| \otimes \hat{I})|A_\sigma\rangle \otimes |v_\sigma\rangle\phi^\sigma \tag{63}$$

where $C$ is an unimportant normalization constant. Because of the post-selection, the system is now again in a non-entangled state so that the partial trace of $|\Psi^{(f)}\rangle\langle\Psi^{(f)}|$ over the first subsystem gives us $|\phi^{(f)}\rangle = C\langle\beta|A_\sigma\rangle\, |v_\sigma\rangle\phi^\sigma$. Making the following phase choices $\langle\beta|A_0\rangle = |\langle\beta|A_0\rangle| e^{i\beta_0}$ and $\langle\beta|A_1\rangle = |\langle\beta|A_1\rangle| e^{-i\beta_1}$, we can again compute the probability of finding the second subsystem in state $|\pi/2,0\rangle$ and again one finds that for a fixed angle $\theta$, the maximum probability occurs for $\varphi' = \beta_0 + \beta_1 = arg\,(\langle\beta|A_0\rangle\langle A^1|\beta\rangle)$. This implies that there is an overall phase change $\Theta$ given by $\Theta = \varphi' - \varphi = arg(\langle A^1|\beta\rangle\langle\beta|A_0\rangle\langle A^0|A_1\rangle)$ which is a well-known *geometric invariant* in the sense

that it depends only on the projection of the state-vectors $|A_0\rangle$, $|A_1\rangle$ and $|\beta\rangle$ on $\mathbb{C}P(N)$. In fact, this quantity is the intrinsic geometric phase picked by a state-vector that is parallel transported through the closed geodesic triangle defined by the projection of the three states on the space of rays. For a single qubit, the geometric invariant is proportional to the area of the geodesic triangle formed by the projection of the kets ($|A_0\rangle$, $|A_1\rangle$ and $|\beta\rangle$) on the Bloch sphere and it is well known to be given by

$$\Theta = arg(\langle A^0|\beta\rangle\langle\beta|A_1\rangle\langle A^1|A_0\rangle) = -\frac{1}{2}\Omega \tag{64}$$

where $\Omega$ is the oriented solid angle formed by the geodesic triangle.

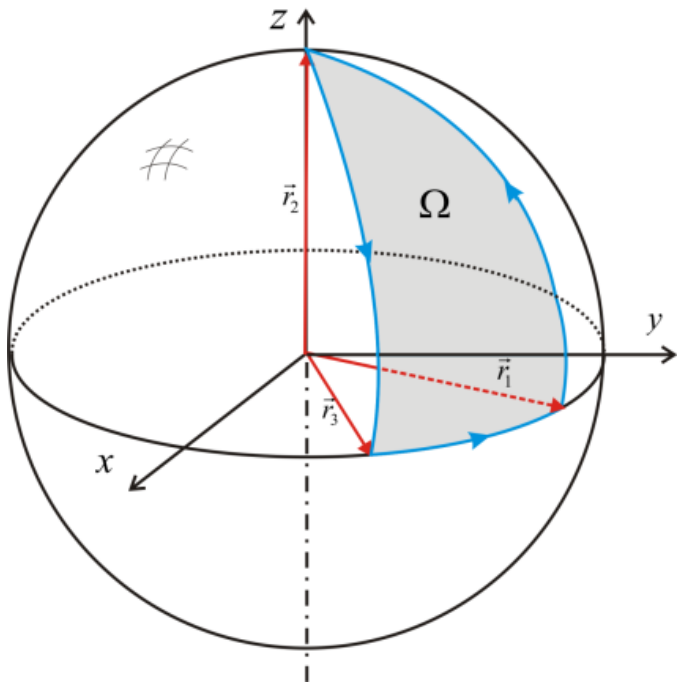

**Figure 2 - Solid angle determined by 3 points: the North Pole and 2 points on the equator**

Returning to the single qubit case, notice that if we choose the following state $|\alpha\rangle = |u_0\rangle$ for the pre-selected state (the "north pole" of the Bloch sphere), and for the post-selected states and $\hat{O} = \hat{\sigma}_1 = |u_0\rangle\langle u^1| + |u_1\rangle\langle u^0|$ as the observable, then it is straightforward to compute the weak value as $O_w = tan\,(\theta/2)e^{i\varphi}$ which is clearly complex-valued in general. What is curious about this result is that the weak value gives a direct physical meaning to the complex projective coordinate of the state vector in the Bloch Sphere.

Suppose now that the physical system $\boldsymbol{W}$ is composed by two subsystems $\boldsymbol{W}^{(S)}\otimes\boldsymbol{W}^{(\infty)}$ as before, but the measuring system $\boldsymbol{W}^{(\infty)}$ is spanned by a complete basis of position kets $\{|q(x)\rangle\}$ (momentum kets $\{|p(y)\rangle\}$), with $-\infty < x, y < +\infty$. And again let us consider the initial state as the product state vector $|\Psi^{(i)}\rangle = |\alpha\rangle\otimes|\phi^{(i)}\rangle$ together with an instantaneous interaction coupling the observable $\hat{O}$of $\boldsymbol{W}^{(S)}$with the momentum observable $\hat{P}$ in $\boldsymbol{W}^{(\infty)}$.The system evolves then to

$$|\Psi^{(f)}\rangle = \int_{-\infty}^{+\infty} dy|A(y)\rangle \otimes |p(y)\rangle\phi_p(y)$$

where $|A(y)\rangle = e^{-i\lambda y\hat{O}}|\alpha\rangle$ is the continuous indexed states that are correlated to the momentum basis of the measuring apparatus and $\phi_p(y) = \langle p(y)|\phi^{(i)}\rangle$ is the wave function of the initial state of the apparatus in the momentum basis. We may now compute (to first order in $dy$) the *intrinsic* phase shift between $|A(y)\rangle$ and $|A(y+dy)\rangle$ in a similar manner that was carried out before with the discretely parameterized states:

$$arg(\langle A(y)|A(y+dy)\rangle \approx -\lambda dy\langle\hat{O}\rangle_{|\alpha\rangle} \tag{65}$$

where $\langle\hat{O}\rangle_{|\alpha\rangle} = \langle\alpha|\hat{O}|\alpha\rangle$ is the expectation value of observable $\hat{O}$ in the initial state $|\alpha\rangle$.We can also compute the shift of the expectation value of the position observable of the particle of the measuring system between the initial and final states. Let $\{|o_j\rangle\}$ $(j = 0,1,\dots N-1)$ be a complete set of eigenkets of observable $\hat{O}$. The final state of the composite system can be described by the following pure density matrix:

$$\hat{\rho}_{|\Psi^{(f)}\rangle} = |\Psi^{(f)}\rangle\langle\Psi^{(f)}| = |o_j\rangle\langle o^k| \otimes \alpha^j \hat{V}^{\dagger}_{\lambda o_j} |\phi^{(i)}\rangle\langle\phi^{(i)}| \hat{V}_{\lambda o_k} \bar{\alpha}_k \tag{66}$$

By taking the partial trace of the $\boldsymbol{W}^{(S)}$ system, we arrive at the following mixed state that describes the measuring system at instant $t_f$:

$$\hat{\rho}^{(M)}_{|\Psi^{(f)}\rangle} = \sum_j |\alpha_j|^2 \hat{V}^{\dagger}_{\lambda o_j} |\phi^{(i)}\rangle\langle\phi^{(i)}| \hat{V}_{\lambda o_j} \tag{67}$$

The ensemble expectation value of position is then

$$[\hat{Q}]_{\hat{\rho}^{(M)}_{|\Psi^{(f)}\rangle}} = tr\left(\hat{\rho}^{(M)}_{|\Psi^{(f)}\rangle}\hat{Q}\right) = \langle\hat{Q}\rangle_{|\phi^{(i)}\rangle} + \lambda\langle\hat{O}\rangle_{|\alpha\rangle} \tag{68}$$

A geometric interpretation of this von Neumann's pre-measurement can be presented in the following way: Let $|\psi(t)\rangle$ be the curve of normalized state vectors in $\boldsymbol{W}^{(N+1)}$ given by the unitary evolution generated by a given hamiltonian $\hat{H}(t)$. The Schrödinger equation implies a relation between $|\psi(t)\rangle$ and $|\psi(t+dt)\rangle$ given by:

$$d|\psi(t)\rangle = |\psi(t+dt)\rangle - |\psi(t)\rangle = -i\hat{H}|\psi(t)\rangle dt \tag{69}$$

The above equation together with (40) lead to a very elegant relation for the squared distance between two infinitesimally nearby projection of state vectors connected by the unitary evolution over $\mathbb{C}P(N)$[26]:

$$ds^2(\mathbb{C}P(N)) = \left[\langle\psi(t)|\hat{H}^2|\psi(t)\rangle - \langle\psi(t)|\hat{H}|\psi(t)\rangle^2\right]dt^2 = \delta^2_{|\psi(t)\rangle}E \tag{70}$$

The above equation means that the speed of the projection over $\mathbb{C}P(N)$ equals the *instantaneous* energy uncertainty $\frac{ds}{dt} = \delta_{|\psi(t)\rangle}E$. A beautiful geometric derivation of the time-energy uncertainty relation that follows directly from this equation can be found in [26,27].

Back to our discussion of the interaction between the systems $\boldsymbol{W}^{(S)}$and $\boldsymbol{W}^{(\infty)}$, note that $|A(y)\rangle = e^{-i\lambda y\hat{O}}|\alpha\rangle$ is formally equivalent to the unitary time evolution equation $|\psi(t)\rangle = e^{-i\hat{H}t}|\psi(0)\rangle$ which is clearly a solution of the Schrödinger equation with a time-independent Hamiltonian. A formal analogy between the two distinct physical processes is exemplified by the association below:

| | |
|---|---|
| $\|\psi(t)\rangle$ | $\|A(y)\rangle$ |
| $\|\psi(0)\rangle$ | $\|\alpha\rangle = \|A(0)\rangle$ |
| $t$ | $y$ |
| $\hat{H}$ | $\lambda\hat{O}$ |

Looking at subsystem $\boldsymbol{W}^{(S)}$and regarding $y$ as an external parameter (just like the time variable for the unitary time evolution) we may write the analog of (70) in $\mathbb{C}P(N) \subset \boldsymbol{W}^{(S)}$:

$$ds^2 = \left[\langle A(y)|\hat{O}^2|A(y)\rangle - \langle A(y)|\hat{O}|A(y)\rangle^2\right]\lambda^2 dy^2 = \left[\langle\alpha|\hat{O}^2|\alpha\rangle - \langle\alpha|\hat{O}|\alpha\rangle^2\right]\lambda^2 dy^2.$$

Comparing this result with equations (40) and (65) we can immediately see the geometric interpretation for the expectation value $\langle\alpha|\hat{O}|\alpha\rangle$ in terms of the $U(1)$ fiber bundle structure as one can easily infer from the pictorial representation in the Figure3 below.

For the case of a *weak measurement*, we choose again the Hamiltonian given by equation (50). Given the initial unentangled state $|\Psi^{(i)}\rangle = |\alpha\rangle \otimes |\phi^{(i)}\rangle$ at $t_0$, the evolution of the system is described as

$$|\Psi^{(f)}\rangle = \int_{-\infty}^{+\infty} dy |A(y)\rangle \otimes |p(y)\rangle \phi_p(y) \tag{71}$$

with $|A(y)\rangle = e^{-i\epsilon y\hat{O}}|\alpha\rangle$.

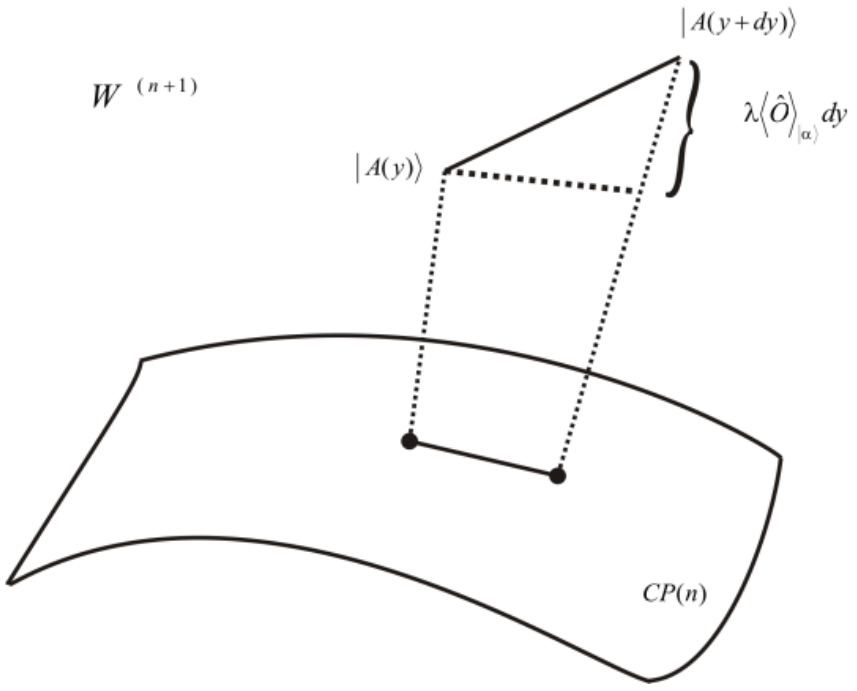

**Figure3- Pictorial representation of the phase difference between $|A(y)\rangle$ and $|A(y+dy)\rangle$**

The global geometric phase related to the infinitesimal geodesic triangle formed by the projections of $|A(y)\rangle$, $|A(y+dy)\rangle$ and the post-selected state $|\beta\rangle$ on $\mathbb{C}P(N)$ (see the Figure 4) is clearly given by

$$\Theta = arg(\langle A(y)|\beta\rangle\langle\beta|A(y+dy)\rangle\langle A(y+dy)|A(y)\rangle) \tag{72}$$

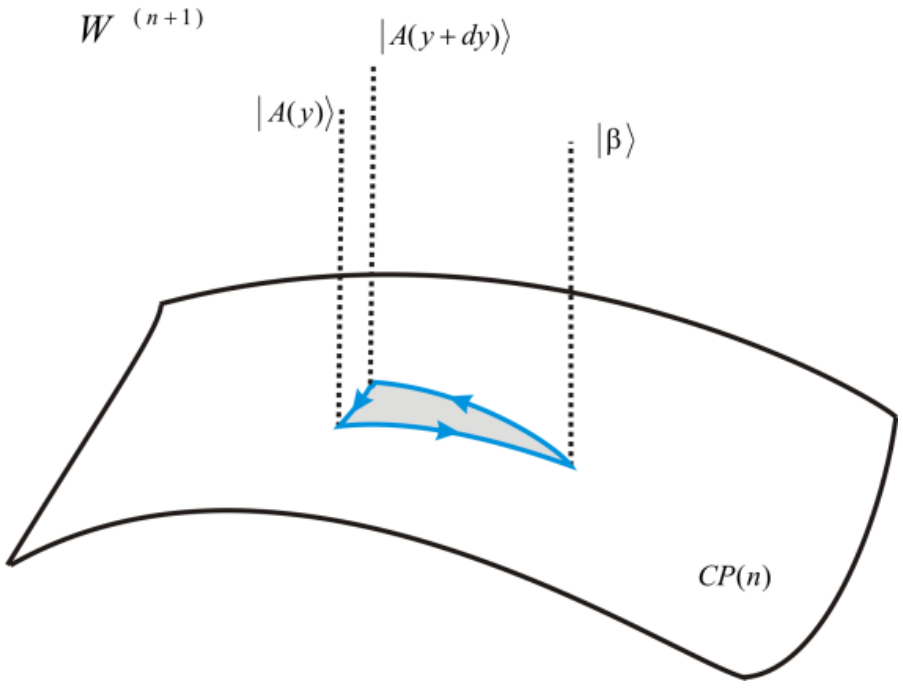

**Figure 4 - Pictorial representation of the global geometric phase**

Expanding to first order in $\epsilon$, we obtain

$$\Theta = -\epsilon\left[Re(O_w) - \langle\hat{O}\rangle_{|\alpha\rangle}\right]dy \tag{73}$$

The final state after post-selection of a state $|\beta\rangle$ of system $\boldsymbol{W}^{(S)}$ is given by

$$|\Psi^{(f)}\rangle \approx C\big(|\beta\rangle\langle\beta| \otimes \hat{I}\big)\big(\hat{I} - i\epsilon\hat{O} \otimes \hat{P}\big)|\alpha\rangle \otimes |\phi^{(i)}\rangle \tag{74}$$

where

$$C \approx \frac{1 + \epsilon\langle P\rangle_{|\alpha\rangle} Im(O_w)}{|\langle\beta|\alpha\rangle|} \tag{75}$$

is the normalization constant because, in general, the state after post-selection is not normalized. By partial tracing out the first subsystem we arrive at:

$$\hat{\rho}_{|\Psi^{(f)}\rangle} = tr_1\left(|\Psi^{(f)}\rangle\langle\Psi^{(f)}|\right)$$
$$= \left[1 - i\epsilon\langle\hat{P}\rangle_{|\phi^{(i)}\rangle}(O_w - \bar{O}_w)\right]|\phi^{(i)}\rangle\langle\phi^{(i)}| - i\epsilon\left(O_w\hat{P}|\phi^{(i)}\rangle\langle\phi^{(i)}| - \bar{O}_w|\phi^{(i)}\rangle\langle\phi^{(i)}|\hat{P}\right)$$

The shift in the ensemble average can then be easily computed as

$$\Delta\hat{Q} = [\hat{Q}]_{\hat{\rho}_{|\Psi^{(f)}\rangle}} - [\hat{Q}]_{\hat{\rho}_{|\Psi^{(i)}\rangle}}$$
$$= \epsilon\left[Im(O_w)\left(\langle\phi^{(i)}|\{\hat{Q},\hat{P}\}|\phi^{(i)}\rangle - 2\langle\hat{P}\rangle_{|\phi^{(i)}\rangle}\langle\hat{Q}\rangle_{|\phi^{(i)}\rangle}\right) + Re(O_w)\right]$$

which is essentially equation (53) again.

# 7 - The Finite Phase Space

## *7.1 Schwinger's Quantum Kinematics:*

(The main reference for this subsection is [11,28])

Let $\boldsymbol{W}^{(N)}$ be an $N$-dimensional quantum space generated by an orthonormal basis $\{|u_k\rangle\}(k = 0,1,2,\ldots N-1)$ which means that $|u_k\rangle\langle u^k| = \hat{I}$ and $\langle u^j|u_k\rangle = \delta_k^j$. These are considered to be *finite position* states. We also define a unitary translation operator $\hat{V}$ that acts cyclically over this basis:

$$\hat{V}|u_k\rangle = |u_{k-1}\rangle \tag{76}$$

The cyclicity means that the following periodic boundary condition must be obeyed:

$$|u_{k+N}\rangle = |u_k\rangle \tag{77}$$

Which implies that $(\hat{V})^N = \hat{I}$ so that the eigenvalues of $\hat{V}$ consists of the $N^{\text{th}}$ roots of unity:

$$\hat{V}|v_k\rangle = v_k|v_k\rangle \quad \text{with} \quad (v_k)^N = 1 \tag{78}$$

The set $\{|v_k\rangle\}, (k = 0,1,2,\ldots N-1)$ is also an orthonormal basis of $\boldsymbol{W}^{(N)}$(the finite set of momentum states) and the $N$distinct eigenvalues are explicitly given by

$$v_k = e^{\frac{2\pi i}{N}k} \tag{79}$$

With a convenient choice of phase, one can show that $\langle u^j|u_k\rangle = \frac{1}{\sqrt{N}}e^{\frac{2\pi i}{N}jk} = \frac{1}{\sqrt{N}}v_k^j$ which is a finite analog of the plane wave function. It is not difficult to convince oneself of the following property:

$$\frac{1}{N}\sum_{l=0}^{N-1} e^{\frac{2\pi i}{N}(j-k)l} = \delta_{jk} \tag{80}$$

One may then define a unitary translator $\hat{U}$ that acts cyclically upon the momentum basis analogously as was carried out for the $\hat{V}$ operator:

$$\hat{U}|v_k\rangle = |v_{k+1}\rangle \tag{81}$$

The same analysis applies now to the $\hat{U}$ operator and amazingly, the eigenstates of $\hat{U}$ can be shown to coincide with the original finite set of position basis with the same spectrum of $\hat{V}$:

$$\hat{U}|u_k\rangle = v_k|u_k\rangle \quad (82)$$

The finite index set $j, k = 0,1,2,\dots,N-1$ takes values in the finite ring $Z_N$ of $mod\, N$ integers. When $N = p$ is a *prime* number, $Z_p$ has the structure of a *finite field*. One distinguished property that is not difficult to derive is the well-known Weyl commutation relations between powers of the unitary translator operators:

$$\hat{V}^j\hat{U}^k = v^{jk}\hat{U}^k\hat{V}^j \quad (83)$$

The above equation is the finite analog of (10). We can also define the *Finite* or *Discrete Fourier Transform* $\hat{F}$ as

$$\hat{F} = |v_j\rangle\langle u^j| \quad (84)$$

The unitarity of $\hat{F}$ is evident since, by construction, it takes one orthonormal basis (the position basis) to another (the momentum basis). One remarkable property is that in both basis, the matrix of $\hat{F}$ is given by

$$F_k^j = \langle u^j|v_k\rangle = \frac{1}{\sqrt{N}}v_k^j \quad (85)$$

### *7.2 The Finite Weyl-Wigner Formalism*

Let us suppose that the dimension $N$ of a finite dimensional space $\boldsymbol{W^{(N)}}$ is an *odd* number. We can define a complete set of $N^2$operators $\hat{\Delta}_{mn}$ analogously as was carried out for the infinitely continuous case:

$$\hat{\Delta}_{mn} = \hat{V}^{-n}\hat{U}^{2m}\hat{V}^{-n}\hat{F}^2, \qquad (m, n \in Z_N) \quad (86)$$

It can be shown also that these finite analogs obey similar properties as their continuous counterparts. They form a complete hermitian and unit-trace set of operators:

$$\left(\hat{\Delta}_{mn}\right)^\dagger = \hat{\Delta}_{mn} \qquad \textit{(hermiticity)}$$

$$tr\left(\hat{\Delta}_{mn}\right) = 1 \qquad \textit{(unit-trace)}$$

The completeness can be written as

$$\sum_{m,n\,\in\, Z_N} \hat{\Delta}_{mn}\hat{O}\hat{\Delta}_{mn} = Ntr\left(\hat{O}\right)\hat{I} \quad (87)$$

for all operators $\hat{O}$. The operator basis $\hat{\Delta}_{mn}$ can thus be *normalized* (within the Hilbert-Schmidt inner product) by a factor of $1/\sqrt{N}$. The completeness and orthonormality of the operator basis can also be expressed in a similar manner that was accomplished within the extended Dirac bra and ket notation introduced in section 2 for the continuous case:

$$|\hat{\Delta}_{mn})(\hat{\Delta}^{mn}| = N\hat{I} \quad \text{and} \quad \left(\hat{\Delta}^{mn}\middle|\hat{\Delta}_{pq}\right) = N\delta_p^m\delta_q^n \quad (88)$$

where $\hat{\Delta}^{mn} = \left(\hat{\Delta}_{mn}\right)^\dagger = \hat{\Delta}_{mn}$. It is now straightforward to define an analogous Weyl-Wigner transform at least for *odd-dimensional* spaces. Given an arbitrary observable $\hat{O}$, its transform is given by the inner product

$$O^{mn} = \left(\hat{\Delta}^{mn}\middle|\hat{O}\right) = tr(\hat{\Delta}_{mn}\hat{O}) \quad (89)$$

An arbitrary operator can clearly be expanded in such basis with a very compact notation:

$$\hat{O} = \frac{1}{N}|\hat{\Delta}_{mn})\left(\hat{\Delta}^{mn}\middle|\hat{O}\right) = \frac{1}{N}\hat{\Delta}_{mn}O^{mn} \quad (90)$$

The inner product between two distinct operators $\hat{A}$ and $\hat{B}$ can be easily computed from their transforms as

$$(\hat{A}|\hat{B}) = \frac{1}{N}\bar{a}_{mn}b^{mn} \tag{91}$$

The $N^2$ operators of the finite Weyl-Wigner (WW) basis (for odd $N$) clearly span the algebra of the Lie Group $U(N)$, even though the basis that is usually chosen to present $u(N)$ is given by the *direct sum* of the $N^2 - 1$ generators of $SU(N)$ plus the identity operator (which generates the trivial $U(1)$ phase). The WW operator basis can be thought as a "rotated" basis that "scrambles up" this distinction as each $\widehat{\Delta}_{mn}$ has a *non-null* component in the direction of the identity operator. Yet, the WW basis has a unique feature. Each element of the basis obeys the following remarkable property

$$\left(\widehat{\Delta}_{mn}\right)^2 = \hat{I} \ (m, n \in Z_N) \tag{92}$$

which makes it easy to be exponentiated, obtaining one-parameter curves in $U(N)$:

$$e^{i\lambda\widehat{\Delta}_{mn}} = \hat{I}\, cos(\lambda) + i\widehat{\Delta}_{mn} sin\,(\lambda) \tag{93}$$

The product of two elements of the basis can be written in a simple way as

$$\widehat{\Delta}_{mn}\widehat{\Delta}_{pq} = v_{2(mp-nq)}\widehat{\Delta}_{m-p,n-q}\hat{F}^2 \tag{94}$$

Notice the "symplectic phase" that appears in the above equation. This result can be further expanded in the WW delta basis because of the following double Fourier identity:

$$\widehat{\Delta}_{pq}\hat{F}^2 = \frac{1}{N}\widehat{\Delta}_{ks}v_q^{-2k}v_p^{-2s} \tag{95}$$

which also allows us to write in a very elegant way the *structure constants* of the algebra $u(N)$.Let us establish first a more convenient "vector like" notation for the points in the "finite or discrete phase space" $Z_N \times Z_N$ by setting:

$$\vec{a} = \binom{m}{n},\ \vec{b} = \binom{p}{q} \text{ and } \vec{c} = \binom{r}{s}$$

for these finite phase plane vectors and the "finite symplectic area" $\Omega(\vec{a}, \vec{b}) = pn - mq$ defined by the vectors $\vec{a}$ and $\vec{b}$ and also by defining the *symbol* $\{\vec{a}, \vec{b}, \vec{c}\}$ over $(Z_N \times Z_N)^3$ with values in $Z_N$:

$$\{\vec{a}, \vec{b}, \vec{c}\} = 2[\Omega(\vec{a}, \vec{b}) + \Omega(\vec{b}, \vec{c}) + \Omega(\vec{c}, \vec{a})] \tag{96}$$

(Note that the *symbol* of three vectors in phase space vanishes whenever *at least two* of them coincide). The Lie algebra $u(N)$ can then be expressed as

$$[\widehat{\Delta}_{\vec{a}}, \widehat{\Delta}_{\vec{b}}] = \tfrac{2\pi}{N}\widehat{\Delta}_{\vec{c}}\Lambda_{\vec{a}\vec{b}}^{\vec{c}} \qquad \text{(sum over } \vec{c}\text{)} \tag{97}$$

with the structure constants $\Lambda_{\vec{a}\vec{b}}^{\vec{c}}$ written elegantly in the following compact manner:

$$\Lambda_{\vec{a}\vec{b}}^{\vec{c}} = Im\left(v_{\{\vec{a},\vec{b},\vec{c}\}}\right) = sin\left[\left(\frac{2\pi}{N}\right)\{\vec{a}, \vec{b}, \vec{c}\}\right] \tag{98}$$

There is ***no** way* to write an *intrinsic formula* for a Weyl-Wigner delta basis for *even-dimensional* spaces. This is because as we have seen above, one must give a meaning to complex phases as $v_{1/2\,mn}$. For odd-dimensional spaces, there is no problem with this because $1/2\, mn$ means $2^{-1}mn$ where the element 2 always has a *multiplicative inverse* in the Ring $Z_N$ for *odd* $N$, but makes ***no** sense* for *even* $N$. This does **not** mean, however, that one *cannot* define a delta WW basis for even-dimensional spaces. One needs only to take special care because of the lack of $mod\, N$ invariance which must be "reinstated by hand". We shall agree that what we *mean* when we write the phase $v_{1/2 mn} = e^{i\frac{\pi mn}{N}}$ makes sense *only* when the representatives of $m, n \in Z \; of \; Z_N$ are taken between $0 \le m, n \le N-1$. With this *precaution*, one can define a WW basis as the following double sum:

$$\hat{\Delta}_{pq} = \frac{v_{\frac{1}{2}mn}}{N} v_{-mp} v_{-nq} \hat{U}^m \hat{V}^n \tag{99}$$

This leads to a complete hermitian orthonormal basis with unit trace as before, but what is *lost* is the nice property that the *square* of the *operators* are *proportional* to the *identity operator* as one can easily check by computing below the $N = 2$ case:

$$\hat{\Delta}_{00} = \frac{1}{2}\begin{pmatrix} 2 & 1+i \\ 1-i & 0 \end{pmatrix}, \qquad \hat{\Delta}_{01} = \frac{1}{2}\begin{pmatrix} 0 & 1-i \\ 1+i & 2 \end{pmatrix},$$
$$\hat{\Delta}_{10} = \frac{1}{2}\begin{pmatrix} 2 & -1-i \\ -1+i & 0 \end{pmatrix}, \qquad \hat{\Delta}_{11} = \frac{1}{2}\begin{pmatrix} 0 & -1+i \\ -1-i & 2 \end{pmatrix}$$

This is only a first example of a certain *anomalous* behavior of *even-dimensional* spaces when one attempts to look for discrete analogs of well-known algebraic properties of the continuous case.

### *7.3 - Finite Coherent States and the choice of a reference state*

(The main references for this section are [29],[30],[31], [32] and [33])

By proceeding with this program of searching for *discrete analogs* of the continuous case, it seems natural to look now for a proper definition of *finite coherent states*. We start by defining a discrete analog of the *displacement operator*:

$$\hat{D}_{mn} = v^{-\frac{mn}{2}} \hat{U}^m \hat{V}^{-n} \tag{100}$$

where $v^{-\frac{mn}{2}}$ is the phase discussed in the previous section. We are now tempted to define $N^2$ coherent states as

$$|m, n\rangle = \hat{D}_{mn}|0\rangle \tag{101}$$

where $|0\rangle$ is some "convenient state of reference" analogous to the ground state of the harmonic oscillator for the continuous case. But, unfortunately, as far as we know, *no* "natural way" to define such state has been presented so far in the literature, even though a number of suggestions have been put forward [29,30,31,32,33,34,35,36,37,38] Since the vacuum state is the *lowest* energy level of the infinite enumerable set $|0\rangle, |1\rangle, |2\rangle, \ldots$ which are *also* eigenstates of the Fourier transform operator, one could envisage a finite analog spectrum of states $|0\rangle, |1\rangle, |2\rangle, \ldots, |N-1\rangle$ with the *correct* eigenvalues of the finite Fourier transform. But this naïve approach is simply *not* possible because of a fact that was known already by Gauss. One should expect then that the finite set of $N$ states above should also be eigenstates of $\hat{F}_{(N)}$ with a set of eigenvalues given by $(i)^0, (i)^1, \ldots, (i)^{N-1}$. This should imply that the *trace* of the Fourier transform should be given by

$$\sum_{m=0}^{N-1} (i)^m = \frac{1-(i)^N}{1-i} \tag{102}$$

But it is clear that the trace of the finite Fourier transform can be computed as the following celebrated *Gaussian sum*:

$$tr(\hat{F}_{(N)}) = \frac{1}{N} \sum_{m=0}^{N-1} e^{\frac{2\pi i}{N} m^2} \tag{103}$$

The result (which is not trivial at all to calculate) can be shown to be $\frac{1-(i)^N}{1-i}$. Gauss himself gave many proofs of this remarkable fact any many other proofs where provided since then by a sequence of distinguished number theorists. Note that these two results are different in general, but for *odd N* they *do coincide*. It would be interesting to construct in a "natural manner" some finite analogs of the *number operator* and/or the *creation* and *annihilation operators* at least for odd dimensional spaces. This would add a somewhat "physical" or "natural" motivated proof to the long list of known proofs of this amazing mathematical fact. One possible choice for a state that is an eigenstate of the finite Fourier transform (with *null* "eigenphase" as is the case for the ground-state of the harmonic oscillator) is the following:

$$|0\rangle = \frac{|u_0\rangle + |v_0\rangle}{\||u_0\rangle + |v_0\rangle\|} \tag{104}$$

We do not claim that this choice is "natural" in any sense, but it clearly obeys the condition $\hat{F}|0\rangle = |0\rangle$ and it is also easy to compute with. Indeed, by defining the finite coherent states with this reference state, it is not difficult to show that

$$\langle p,q|r,s\rangle = v_{\frac{1}{2}(rq-ps)} \times \begin{cases} 1 \\ 1/2(N+2\sqrt{N})/(N+\sqrt{N}) \\ 1/2(N+2\sqrt{N})/(N+\sqrt{N}) \\ \cos\left[\pi(r-p)(s-q)/N\right] \end{cases} \text{if} \begin{cases} r = p \text{ and } q = s \\ r \neq p \text{ and } q = s \\ r = p \text{ and } q \neq s \\ r \neq p \text{ and } q \neq s \end{cases}$$

Note here that we have a *symplectic phase* similar to the continuous case, but for *even-dimensional spaces* (unlike the continuous case) the cosine may become *null* for some values, which means that there exist *distinct* coherent states *orthogonal* to each other, something clearly ruled out for the continuous case. Again we find a somewhat anomalous behavior for even-dimensional spaces.

# 8 - Schwinger's formalism for Finite Quantum Mechanics and Aharonov's Modular Variables

In [39], Aharonov and collaborators introduced the concept of modular variables to explain some peculiar non-local quantum effects as the modular momentum exchange between particles and fields in situations like the well-known Aharonov-Bohm (AB) effect where a beam of electrons suffers a phase shift from the magnetic field of a solenoid even without having had any direct contact with the field. This is contrary to the usual view where the AB phenomenon is explained by a *local* interaction between the particles and field potentials, even if the potentials are considered somewhat *unphysical* because they are defined only up to some Gauge transformation. In this section, we carry out the continuous limit of Schwinger's finite structure in order to present a finite analogue of Aharonov's modular variable concept and we also discuss the concept of *pseudo-degrees of freedom*[40], an idea that is essential to deliver a proper understanding of modular variables.

## *8.1 - The heuristic continuum limit*

The implementation of the "continuum heuristic limit" (when the dimensionality of the quantum spaces approach infinity) can be performed in two distinct manners: one symmetric and the other non-symmetric between the position and momentum states. First, we briefly outline below, the *symmetric* case and following this, we present the *non-symmetric* limit which we shall use to discuss the modular variable concept.

### *8.1.1 - The symmetric continuum limit*

Let the dimension $N$ of the quantum space be an *odd* number (with no loss of generality) and let us *rescale* in equal footing the finite position and momentum states as

$$|q(x_j)\rangle = \left(\frac{N}{2\pi}\right)^{1/4} |u_j\rangle \qquad \text{and} \qquad |p(y_k)\rangle = \left(\frac{N}{2\pi}\right)^{1/4} |v_k\rangle \tag{105}$$

with

$$x_j = \left(\frac{2\pi}{N}\right)^{1/2} j \qquad \text{and} \qquad y_k = \left(\frac{2\pi}{N}\right)^{1/2} k \tag{106}$$

so that the discrete indices are disposed symmetrically in relation to *zero*:

$$j,k = -\frac{N-1}{2}, \dots, +\frac{N-1}{2} \qquad \text{and} \qquad \Delta x_j = \Delta y_k = \left(\frac{2\pi}{N}\right)^{1/2} \tag{107}$$

We may write the completeness relations for both basis as

$$\hat{I} = \sum_{j=-\frac{N-1}{2}}^{j=+\frac{N-1}{2}} \left(\frac{2\pi}{N}\right)^{1/2} |q(x_j)\rangle\langle q(x_j)| = \sum_{k=-\frac{N-1}{2}}^{k=+\frac{N-1}{2}} \left(\frac{2\pi}{N}\right)^{1/2} |p(y_k)\rangle\langle p(y_k)| \tag{108}$$

One can give a natural heuristic interpretation of the $N \to \infty$ limit for the above equation as

$$\hat{I} = \int_{-\infty}^{+\infty} dx |q(x)\rangle\langle q(x)| = \int_{-\infty}^{+\infty} dy |p(y)\rangle\langle p(y)| \tag{109}$$

The generalized orthonormalization relations show that the new defined basis is formed by *singular* state-vectors with "infinite norm":

$$\langle q(x_j)|q(x_k)\rangle = \langle p(x_j)|p(x_k)\rangle = \lim_{N\to\infty} \left(\frac{N}{2\pi}\right)^{1/2} \delta_{jk} \xrightarrow[N\to\infty]{} \delta(x_j - x_k) \tag{110}$$

The above equation also serves as a *heuristic-based* definition of the *Dirac delta function* as a continuous limit of the *discrete Kronecker delta*. Yet the overlap of elements from the distinct basis gives us the well-known plane-wave basis as expected:

$$\langle q(x_j)|p(y_k)\rangle = \frac{1}{\sqrt{2\pi}} e^{ix_j y_k} (\hbar = 1) \tag{111}$$

### *8.1.2 - The non-symmetric continuum limit*

We introduce now a *different* scaling for the variables of position and momentum with a given $\xi \in \mathbb{R}$:

$$|q(x_j)\rangle = \left(\frac{N}{\xi}\right)^{1/2} |u_j\rangle \qquad \text{and} \qquad |p(y_k)\rangle = \left(\frac{\xi}{2\pi}\right)^{1/2} |v_j\rangle \tag{112}$$

with

$$x_j = \frac{\xi}{N} j \qquad \text{and} \qquad y_k = \frac{2\pi}{\xi} k \tag{113}$$

so that $\Delta x_j = \xi/N \to 0$ for $N \to \infty$.Only the*position* states become *singular* and the $x_j$ variable takes value in a bounded quasi-continuum set so that the resolution of identity can be written as

$$\hat{I}_\xi = \int_{-\xi/2}^{+\xi/2} dx |q(x)\rangle\langle q(x)| = \frac{2\pi}{\xi} \sum_{k\in\mathbb{Z}} |p(y_k)\rangle\langle p(y_k)| \tag{114}$$

Note that the momentum states continue to be of *finite norm* and their sum is taken over the enumerable but discrete set of integers $k \in \mathbb{Z}$. The identity operator$\hat{I}_\xi$can be thought as the projection operator on the subspace of *periodic functions* with period $\xi$. The overlap between position and momentum states is again given by the usual plane-wave function

$$\langle q(x_j) | p(y_k) \rangle = \frac{1}{\sqrt{2\pi}} e^{i x_j y_k} (\hbar = 1) \tag{115}$$

Note also that we could have *reversed* the above procedure by choosing from the beginning the opposite scaling for the position and momentum states. In this case, the *momentum* eigenkets would form a continuous bounded set of singular state-vectors and the *position* eigenvectors would form an enumerable infinity of finite norm kets.

### *8.2 - Pseudo degrees of freedom*

In the $x^1 - x^2$ plane, one can easily visualize the translations of the ket $|q(\vec{x})\rangle$ acted repeatedly upon with $\hat{V}_{\vec{\xi}}$ as in Figure 5, where the resulting position kets can be represented on a straight line in the plane that contains point $\vec{x}$ but with slope given by the $\vec{\xi}$ direction.

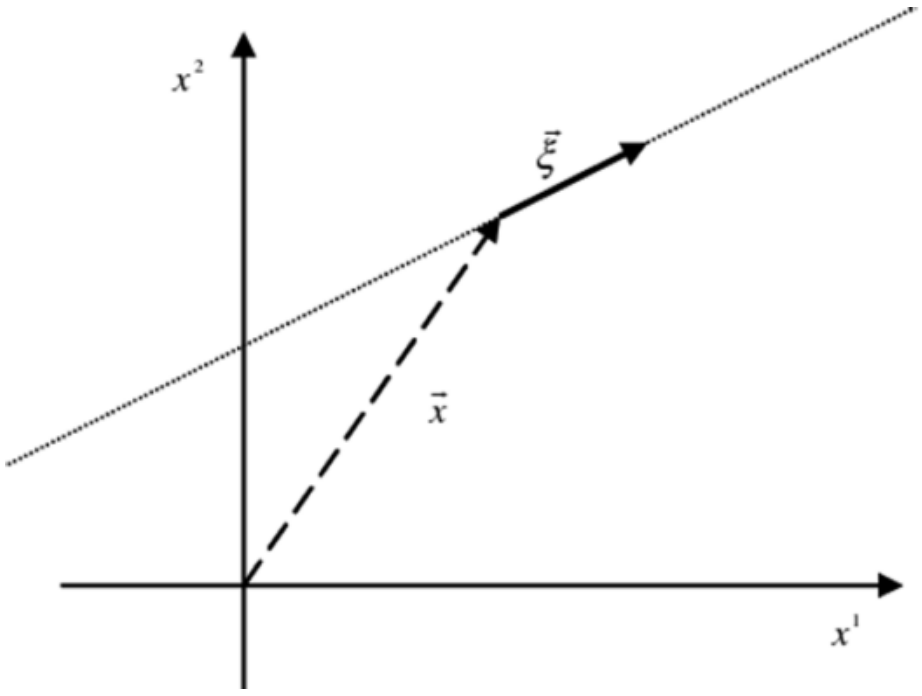

**Figure 5- Two degrees of freedom**

Of course, to reach an arbitrary point in the plane, one needs at least *two* linear independent directions. This is precisely what one means when it is said that the plane is two-dimensional. But things for finite quantum spaces are *not* quite so simple. Let us consider first a 4-dimensional system given by the product of two 2-dimensional spaces (two qubits) $\boldsymbol{W}_{(4)} = \boldsymbol{W}_{(2)} \otimes \boldsymbol{W}_{(2)}$ (it is important to notice here that one must not confuse the dimension of space, the so called degree of freedom with the dimensionality of the quantum vector spaces). We shall discard in the following discussion, the indices that indicate dimensionality to eliminate excessive notation. So let $\{|u_0\rangle, |u_1\rangle\}$ be the position basis for each individual qubit space so that computational (unentangled) basis of the tensor product spaces is $\{|u_0\rangle \otimes |u_0\rangle, |u_0\rangle \otimes |u_1\rangle, |u_1\rangle \otimes |u_0\rangle, |u_1\rangle \otimes |u_1\rangle\}$. One may represent such finite 2-space as the discrete set formed by the four points depicted in the Figure 6:

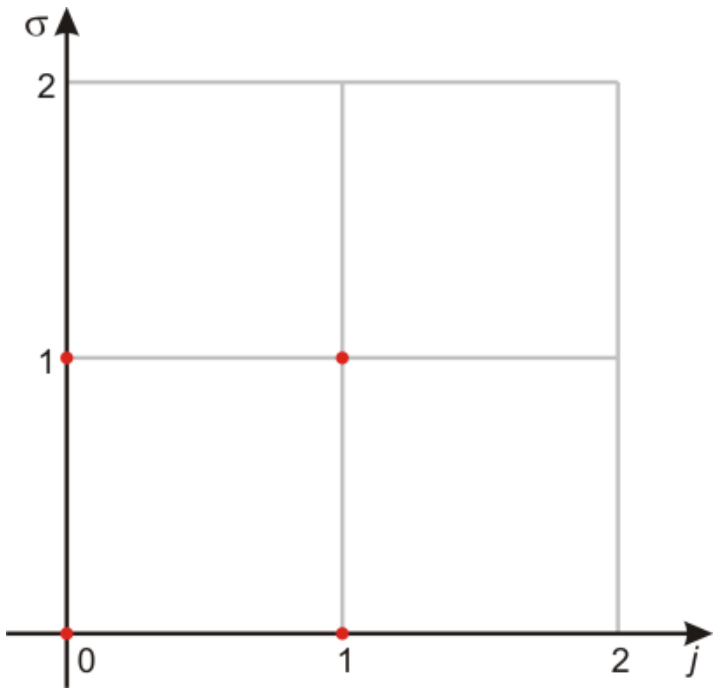

Figure 6 - Finite 2-space for 2 qubits

One may even construct distinct "straight lines" in this discrete two-dimensional space acting upon the computational basis $|u_j\rangle \otimes |u_\sigma\rangle$ with the $\hat{V} \otimes \hat{V}$ operator as shown in the Figure 7.

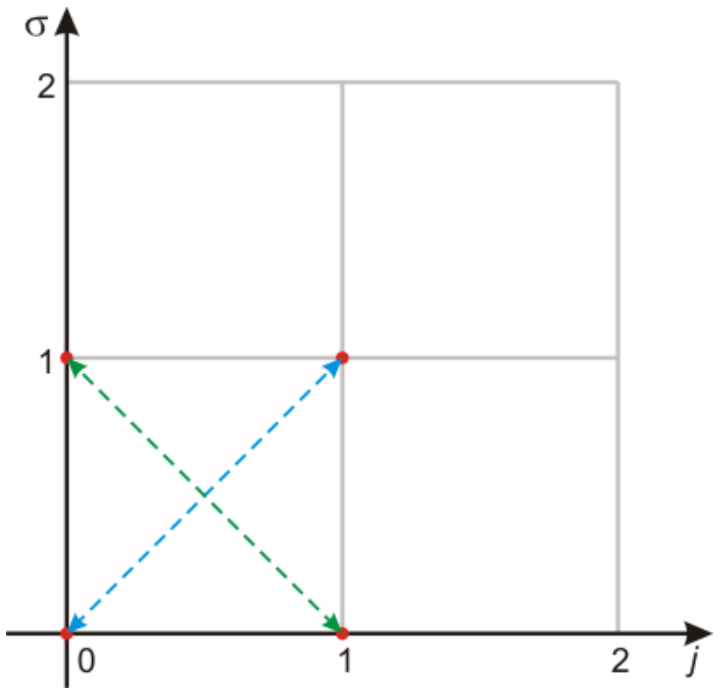

Figure 7 - Discrete parallel lines (0, 0); (1, 1) and (0,1); (1,0)

Each of the two parallel "straight lines" are geometric invariants of the discrete 2-plane under the action of $\hat{V} \otimes \hat{V}$.

Consider now, a six-dimensional quantum space $\boldsymbol{W}_{(6)} = \boldsymbol{W}_{(2)} \otimes \boldsymbol{W}_{(3)}$ given by the product of a qubit and a qutrit with finite position basis given byrespectively $\{|u_0\rangle, |u_1\rangle\}$ and $\{|u_0\rangle, |u_1\rangle, |u_2\rangle\}$. In this case, the fact the dimensions of the individual are *coprime* means that the action of the $\hat{V} \otimes \hat{V}$ operator on the product basis $\{|u_j\rangle \otimes |u_\sigma\rangle\}$, $j = 0{,}1$ and $\sigma = 0{,}1{,}2$ can be identified with the action of $\hat{V}_{(6)} = \hat{V}_{(2)} \otimes \hat{V}_{(3)}$ on the same basis relabeled as $\{|u_0\rangle, |u_1\rangle, |u_2\rangle, |u_3\rangle, |u_4\rangle, |u_5\rangle\}$. One can start with the $|u_0\rangle \otimes |u_0\rangle$ state and cover the *whole space* with *one single line* as shown in the Figure 8.

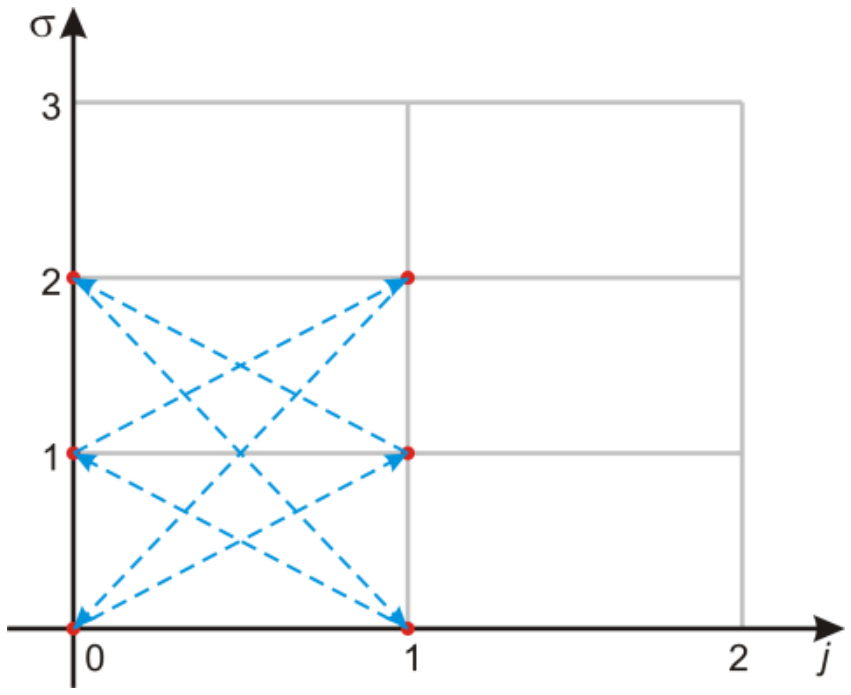

**Figure 8 - A single line covers the whole space**

This reduction of *two* degrees of freedom to *only one* single effective degree of freedom is a general fact for all product spaces when the dimensions of the factor spaces are *co-prime*. This fact follows from elementary number theory and it can be shown that when $MDC(N_a, N_b) = 1$ for quantum spaces $\boldsymbol{W}_{(N_a)}$ and $\boldsymbol{W}_{(N_b)}$ then one may say that the two degrees of freedom are actually *pseudo-degrees of freedom* because one can associate only *one* effective single degree of freedom to the system in this sense. In the case above, it is easy to see that one of the pseudo degrees of freedom is nothing else but a *factorizing period* of the larger space. In this way we can either interpret the above finite six-dimensional position space as three periods of two or as two periods of three. All this is rather obvious and elementary, but surprisingly this is the kind of mathematical structure behind Aharonov's concept of modular variables.

### *8.3 - The N-slit experiment*

Consider the paradigmatic experiment of diffraction of a particle through an apparatus with a large set of equidistant slits as in the Figure 9 below:

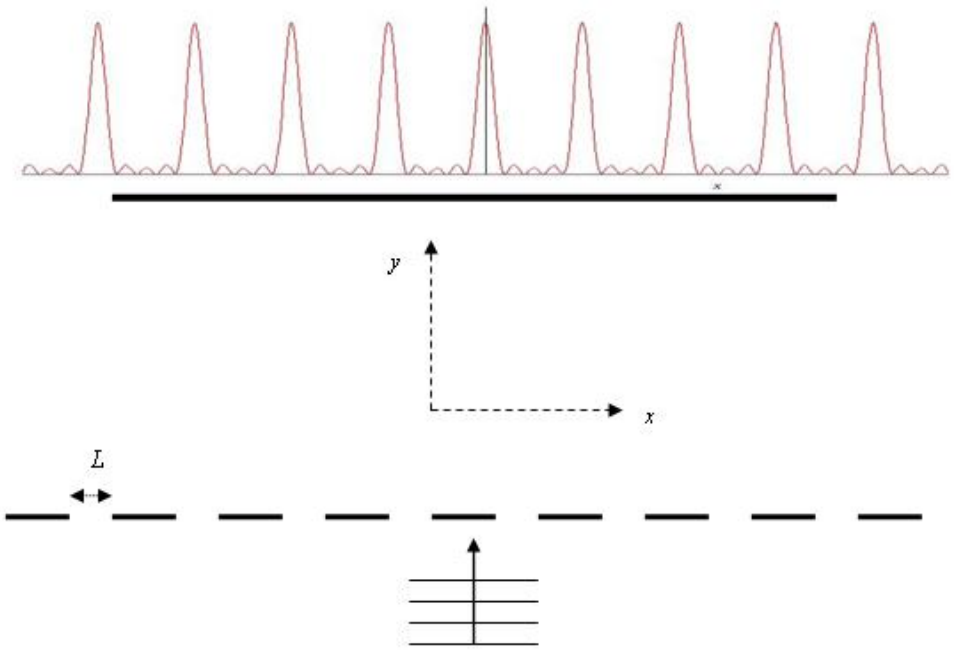

**Figure 9 - n-slit interference experiment**

We can model the interaction of the particle with the slits with the following Hamiltonian in the $x$ direction:

$$\hat{H}(t) = \frac{1}{2m}\hat{P}^2 + V(\hat{Q})\delta(t) \qquad \text{with} \qquad V(\hat{Q} + L\hat{I}) = V(\hat{Q}) \tag{116}$$

where the particle "hits the screen" at $t = 0$. The fundamental physical assumptions here are that the interaction of the particle happens so fast that the unitary time evolution is given by $\hat{U}(t) \approx e^{-iV(\hat{Q})}$. By expanding this function in a Fourier series one gets

$$e^{-iV(\hat{Q})} = \sum_{n\in\mathbb{Z}} c_n e^{\frac{2\pi i n}{L}\hat{Q}} \quad (117)$$

The initial state-vector (in the $x$ direction) of the particle is an eigenstate of momentum with *zero momentum* $|p(0)\rangle$ so the state *just after* the interaction becomes $\hat{U}(t)|p(0)\rangle$. It is not difficult to see that (because of the periodicity of the potential) the coefficients $c_n$ of the expansion *cannot* depend on$n$, so the final state is given by the following *singular* ket

$$c\sum_{n\in\mathbb{Z}} e^{\frac{2\pi i n}{L}\hat{Q}}|p(0)\rangle = c\sum_{n\in\mathbb{Z}} |p(\frac{2\pi n}{L})\rangle$$

The resulting state has the remarkable property of being an eigenstate both of $\hat{U}_{\frac{2\pi}{L}}$ and $\hat{V}_L$.

This mathematical structure is behind the fundamental *dynamical non-locality* involved in the n-slit experiment. Note that because of the Heisenberg-Weyl relation, the $\hat{U}_{2\pi/L}$ and $\hat{V}_L$operators *commute* and since they are unitary, their eigenvalues are necessarily *complex phases*. Aharonov and collaborators named the *phases* of the *simultaneous eigenvalues* of these pairs of operators as *modular variables*. A phase space description of such a state is given by the Figure 10:

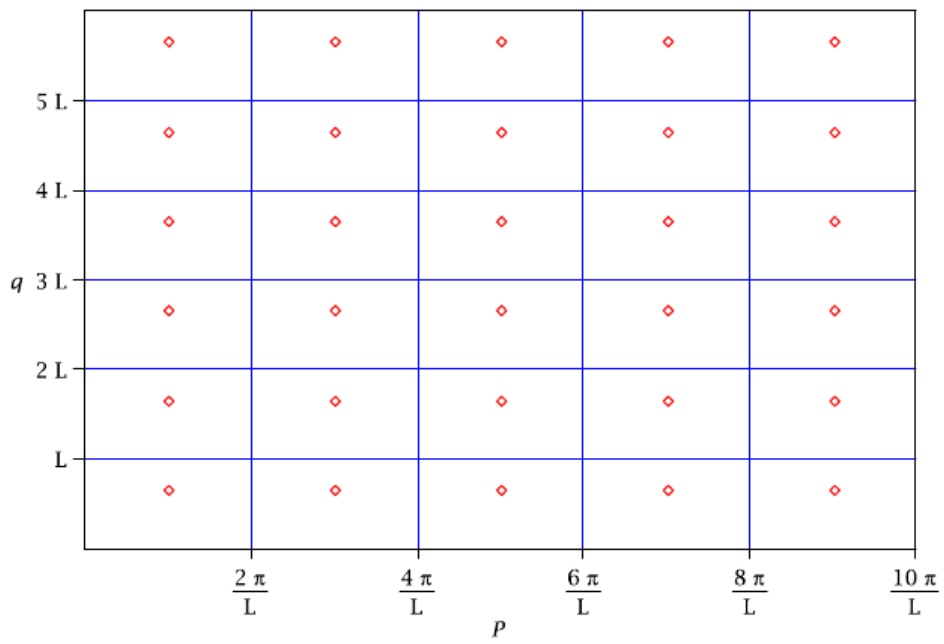

**Figure 10 - state with $q_{mod} = 2L/3$ and $p_{mod} = \pi/L$**

This means that for the state represented above, one has that,in each cell, itis represented by an exact point with sharp values of modular position and momentum but there is a *complete uncertainty* about *which cell* it belongs to. This is a basic feature of the modular variable description. A precise mathematical description of a finite analogue of this phenomenon in terms of the pseudo-degrees of freedom can be given: consider $\boldsymbol{W}_{(N)} = \boldsymbol{W}_{(N_a)} \otimes \boldsymbol{W}_{(N_b)}$ as the state space for a quantum mechanical system with $MDC(N_a, N_b) = 1$. We can then offer an interpretation for this single degree of freedom of $\boldsymbol{W}_{(N)}$ as a degree composed of "$N_b$ periods of size $N_a$" (or vice-versa). In fact, we may define the following state of $\boldsymbol{W}_{(N)}$:

$$|\,j_a, \sigma_b^{(N)}\rangle = |\,v_{j_a}^{(N_a)}\rangle \otimes |\,u_{\sigma_b}^{(N_b)}\rangle \quad (118)$$

This state is simultaneously an eigenstate of *finite momentum* of $\boldsymbol{W}_{(N_a)}$ and finite position of $\boldsymbol{W}_{(N_b)}$ and clearly represents a finite analogue of the state represented in Figure 10. They are also simultaneous eigenstates of the (obviously commuting operators since they act on different spaces) unitary operators $\hat{U}_{(N_a)} \otimes \hat{I}_{(N_b)}$ and $\hat{I}_{(N_a)} \otimes \hat{U}_{(N_b)}$. Let us illustrate this again with an example of our "toy six-dimensional" case:

Let the state $|1,2^{(6)}\rangle = |\,v_1^{(2)}\rangle \otimes |\,u_2^{(3)}\rangle$ be represented by the phase space plot below:

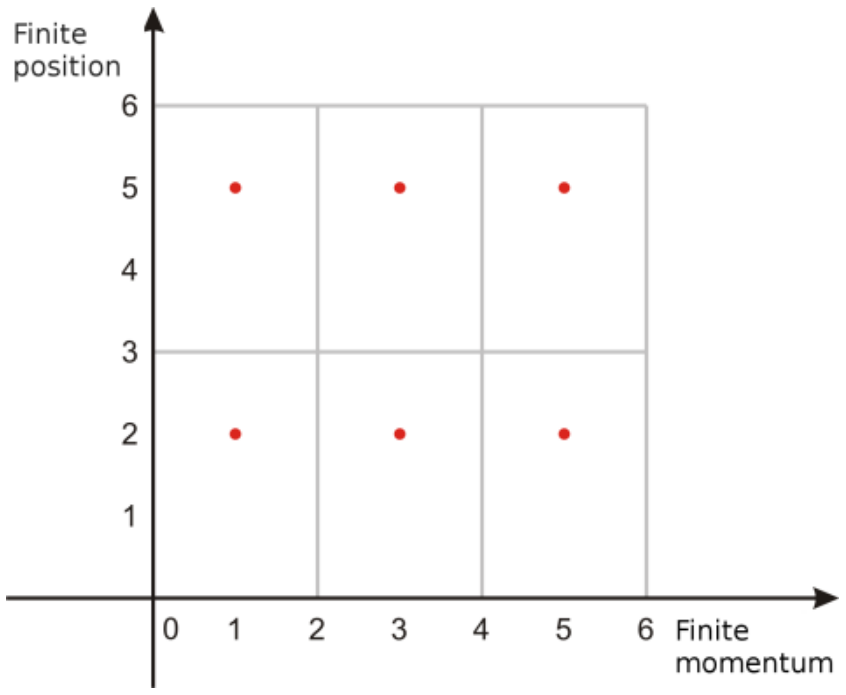

**Figure 11 - state $|1,2^{(6)}\rangle \in W_{(6)}$**

States with this peculiar mathematical structure have been described independently by Zak to study systems with periodic symmetry in quantum mechanics [41,42]. We may call them Aharonov-Zak states (AZ). Theparticular AZ state obtained in the N-slit (with large enough N) can be thought as obtained by an *ideal projective measurement*performed by the slit apparatus onthe second modular variable subspace of the particle:

$$|\, v_0^{(N_a)}\rangle \otimes |\, v_0^{(N_b)}\rangle \rightarrow \underset{\text{in the second subspace}}{\text{projective "space" measurement}} \rightarrow |\, v_0^{(N_a)}\rangle \otimes |\, u_0^{(N_b)}\rangle = \frac{1}{\sqrt{N_b}} \sum_{\sigma_b=0}^{N_b-1} |\, v_0^{(N_a)}\rangle \otimes |\, v_{\sigma_b}^{(N_b)}\rangle$$

The *Continuum limit* of the AZ state can be constructed through the *non-symmetric limit* discussed in a previous subsection. The only care that must be taken is that, given the two subspaces, the *opposite limit* must be taken for each subspace. That is, if one chooses to make the *momentum basis* of the *first subspace* go to infinity as a bounded continuum, then for the *second subspace* it is the *position basis* that must become a bounded continuum set and vice-versa.